\documentclass[floatfix,twocolumn,showpacs,preprintnumbers,amsmath,amssymb,pra,superscriptaddress,longbibliography]{revtex4-1}
\usepackage{color}
\usepackage[usenames,dvipsnames,svgnames,table]{xcolor}
\usepackage[colorlinks=true,linkcolor=blue,urlcolor=blue,citecolor=blue]{hyperref}
\usepackage{mathtools}
\usepackage{graphicx}
\usepackage{dcolumn}
\usepackage{array}
\usepackage{lipsum}
\usepackage{bm}
\usepackage{subfigure}
\usepackage{amssymb}
\usepackage{multirow}
\usepackage{tabularx}
\usepackage{amsmath}
\usepackage{braket}
\usepackage{csquotes}
\graphicspath{{plots/}}
 \usepackage{lipsum}
\usepackage{mathrsfs}
\usepackage{placeins}

\newcommand{\beq}{\begin{equation}}
\newcommand{\eeq}{\end{equation}}
\newcommand{\bea}{\begin{eqnarray}}
\newcommand{\eea}{\end{eqnarray}}

\begin{document}
\title{ Electronic density response of warm dense hydrogen on the nanoscale
}

\author{Tobias Dornheim}
\email{t.dornheim@hzdr.de}

\affiliation{Center for Advanced Systems Understanding (CASUS), D-02826 G\"orlitz, Germany}
\affiliation{Helmholtz-Zentrum Dresden-Rossendorf (HZDR), D-01328 Dresden, Germany}

\author{Maximilian P.~B\"ohme}

\affiliation{Center for Advanced Systems Understanding (CASUS), D-02826 G\"orlitz, Germany}
\affiliation{Helmholtz-Zentrum Dresden-Rossendorf (HZDR), D-01328 Dresden, Germany}
\affiliation{Technische  Universit\"at  Dresden,  D-01062  Dresden,  Germany}

\author{Zhandos~A.~Moldabekov}

\affiliation{Center for Advanced Systems Understanding (CASUS), D-02826 G\"orlitz, Germany}
\affiliation{Helmholtz-Zentrum Dresden-Rossendorf (HZDR), D-01328 Dresden, Germany}



\author{Jan Vorberger}
\affiliation{Helmholtz-Zentrum Dresden-Rossendorf (HZDR), D-01328 Dresden, Germany}



\begin{abstract}
The properties of hydrogen at warm dense matter (WDM) conditions are of high importance for the understanding of astrophysical objects and technological applications such as inertial confinement fusion. In this work, we present extensive new \emph{ab initio} path integral Monte Carlo (PIMC) results for the electronic properties in the Coulomb potential of a fixed ionic configuration. This gives us new insights into the complex interplay between the electronic localization around the protons with their density response to an external harmonic perturbation. We find qualitative agreement between our simulation data and a heuristic model based on the assumption of a local uniform electron gas model, but important trends are not captured by this simplification. In addition to being interesting in their own right, we are convinced that our results will be of high value for future projects, such as the rigorous benchmarking of approximate theories for the simulation of WDM, most notably density functional theory.
\end{abstract}
\maketitle

\section{Introduction\label{sec:introduction}}

Hydrogen constitutes the most abundant element in our universe. Despite its apparent simplicity, it exhibits a plethora of complex and intriguing phenomena, including the metallization transition at high pressure~\cite{Morales_PNAS_2010,Pierleoni_PNAS_2016,Azadi_2013,Knudson_Science_2015,Silvera_Science_2017,Celliers_Science_2018} that might give rise to a potential exotic supersolid state~\cite{Myung_PRL_2022}. Of particular interest are the properties of hydrogen at extreme densities and temperatures. These \emph{warm dense matter} (WDM) conditions are typically defined by two characteristic parameters, that are of the order of one~\cite{wdm_book,Ott2018}: (i) the Wigner-Seitz radius $r_s=\overline{a}/a_\textnormal{B}$ (where $a_\textnormal{B}$ is the Bohr radius), and (ii) the degeneracy temperature $\Theta=k_\textnormal{B}T/E_\textnormal{F}$ (with $E_\textnormal{F}$ being the usual Fermi energy~\cite{quantum_theory}).
In fact, warm dense hydrogen is ubiquitous throughout nature, and occurs in a number of astrophysical objects such as the interior of giant planets~\cite{Benuzzi_Mounaix_2014,drake2018high} and brown dwarfs~\cite{becker,saumon1}.
Moreover, the fuel capsule in inertial confinement fusion (ICF) experiments~\cite{Betti2016} has to traverse the WDM regime~\cite{hu_ICF} on its pathway towards ignition in a controlled way, which makes the accurate understanding of warm dense hydrogen an important step towards the technological utilization of ICF as a source of green energy~\cite{Zylstra_Nature_2022}.

Unfortunately, the theoretical understanding of WDM constitutes a difficult challenge due to the highly nontrivial interplay of various physical effects, including Coulomb coupling between the electrons and ions, quantum degeneracy effects such as diffraction and Pauli blocking, as well as strong thermal excitations out of the ground state~\cite{wdm_book,new_POP}. Indeed, the condition $r_s\sim\Theta\sim1$ that defines the WDM regime often rules out potential expansions around the ground-state limit, or weak-coupling expansions around the noninteracting case such as many-body Green functions~\cite{kremp2005quantum}. This generally makes computational quantum many-body simulation methods the most promising option, with the combination of a classical molecular dynamics (MD) propagation of the ions with the electron-ion forces obtained from density functional theory (DFT) calculations being arguably the work horse of WDM theory.
At ambient conditions, it is known empirically that DFT offers an attractive balance between a manageable computation cost and an often reasonable accuracy for different properties. Extending DFT simulations to extreme conditions is thus associated with two main challenges. 1) It is well known that the computation cost of the standard Kohn-Sham DFT method~\cite{KS65} rapidly increases with the temperature, which constitutes a bottleneck over substantial parts of the WDM regime. To address this obstacle, a number of computationally less expensive methods have been suggested in the literature~\cite{Zhang_POP_2016,BLANCHET2022108215,Ding_PRL_2018,Bethkenhagen_high_T_2021,Fiedler_PRR_2022,White_2022}. 2) The accuracy of any DFT simulation decisively depends on the employed exchange--correlation functional; it cannot be obtained within DFT itself and has to be supplied as an external input.
While the performance of different types of functionals is reasonably well understood at ambient conditions~\cite{Cohen2008Science}, the development of novel functionals that are explicitly designed for applications at WDM conditions is substantially less advanced. At the same time, it has also become clear that the application of ground-state functionals becomes questionable for $\Theta\sim1$~\cite{karasiev_importance,kushal,Sjostrom_PRB_2014,Dharma-wardana_Computation_2016}.

This unsatisfactory situation has started to change only recently with the advent of the first highly accurate parametrizations of the exchange--correlation free energy of a uniform electron gas (UEG)~\cite{review,groth_prl,ksdt,status} based on extensive \emph{ab initio} path integral Monte Carlo (PIMC) simulations~\cite{Brown_PRL_2013,dornheim_prl,Malone_JCP_2015,Malone_PRL_2016,Dornheim_POP_2017,Dornheim_PRB_2016,Groth_PRB_2016,Malone_JCP_2015}. In particular, these parametrizations allow for thermal DFT calculations~\cite{Mermin_DFT_1965} on the level of the local density approximation. Subsequently, Karasiev and co-workers have presented improved functionals~\cite{Karasiev_PRL_2018,Karasiev_PRB_2022} on higher rungs of Jacob's ladder~\cite{Perdew_AIP_2001}. These efforts have been complemented by Moldabekov \emph{et al.}, who have benchmarked different functionals for the weakly nonuniform and the strongly inhomogeneous electron gas \cite{moldabekov2021jcp, moldabekov2022prb, moldabekov2023jcp, moldabekov2023scipost}.
Yet, a rigorous benchmark against exact reference data has hitherto been missing.
We note that this is a general feature of WDM simulations and also applies e.g.~to PIMC simulations based on the de-facto uncontrolled fixed-node approximation~\cite{Ceperley1991,Militzer_PRE_2001}.

Very recently, B\"ohme \emph{et al.}~\cite{Bohme_PRL_2022,Bohme_PRE_2023} have presented the first unbiased PIMC simulation results for the electronic density response of hydrogen in the WDM regime. While being computationally very costly due to the notorious fermion sign problem~\cite{troyer,dornheim_sign_problem}, these calculations did not use any uncontrolled approximations such as the usual restrictions on the nodal surface of the fermionic density matrix. This has allowed them to study different linear-response properties of hydrogen such as the static exchange--correlation kernel---a key property for a multitude of applications~\cite{Moldabekov_JCTC_2023, moldabekov2023jpcl} such as time-dependent DFT calculations~\cite{moldabekov2023linearresponse,Dornheim_review}.
This is particularly important for the modelling and interpretation of X-ray Thomson scattering (XRTS) experiments~\cite{siegfried_review,sheffield2010plasma}, which have emerged as a widely used method of diagnostics for WDM~\cite{Gregori_PRE_2003,kraus_xrts,Dornheim_T_2022,Dornheim_T2_2022,schoerner2023xray,dornheim2023xray}.

In the present work, we substantially extend these efforts by presenting new extensive PIMC simulation data for the electronic density in warm dense hydrogen on the nanoscale. This allows us to study the interplay of the electrons with the protons, and to assess the localization for different parameters. Moreover, we give direct insights into the impact of the protons onto the reaction of the electrons to an external harmonic perturbation, i.e., into the static electronic density response of warm dense hydrogen. 
In addition to being interesting in their own right, our results, having been obtained within the fixed external potential of an ion snapshot, will be of high value for the rigorous benchmarking of thermal DFT simulations of WDM in future works.

The paper is organized as follows. In Sec.~\ref{sec:theory}, we introduce the relevant theoretical background, including a discussion of the relevant system parameters (\ref{sec:Hamiltonian}) and a brief summary of the PIMC simulation set-up (\ref{sec:PIMC}).
Sec.~\ref{sec:results} is devoted to the presentation of our simulation results, starting with an in-depth analysis of the convergence with the number of imaginary-time propagators in Sec.~\ref{sec:convergence}.
We study the electronic localization and static density response to an external harmonic perturbation in the low-density regime ($r_s=4$) in Sec.~\ref{sec:low_density}, and consider the cases of metallic density ($r_s=2$) and high density ($r_s=1$) in the subsequent Sec.~\ref{sec:density}.
The paper is concluded by a summary and outlook in Sec.~\ref{sec:summary}.

\section{Theory\label{sec:theory}}

We assume Hartree atomic units throughout this work.

\subsection{Hamiltonian and system parameters\label{sec:Hamiltonian}}

Following the notation from Ref.~\cite{Bohme_PRE_2023}, we express the Hamiltonian of $N$ electrons (in periodic boundary conditions and a cubic simulation cell of volume $\Omega=L^3$) within the fixed external potential of $N$ protons as
\begin{equation}\label{eq:Hamiltonian}
\hat{H} = \underbrace{-\frac{1}{2} \sum_{l=1}^N \nabla_l^2}_{\hat K} + \underbrace{\hat{W} + \hat{V}_I\left(\{\mathbf{I}_0,\dots,\mathbf{I}_{N-1}\}\right)}_{\hat V}\ ,
\end{equation}
where $\hat{W}$ denotes the electron--electron interaction that we evaluate using the standard Ewald summation technique~\cite{Fraser_PRB_1996}, and $\hat{V}_I$ is the single-particle potential due the ions at positions $\mathbf{I}_0,\dots,\mathbf{I}_{N-1}$. We note that the Hamiltonian (\ref{eq:Hamiltonian}) can be decomposed into a kinetic ($\hat{K}$) and a potential ($\hat{V}$) part, which becomes important for the discussion of the PIMC method in Sec.~\ref{sec:PIMC} below.

To study the electronic density response, we follow Refs.~\cite{moroni,moroni2,bowen2,dornheim_pre,groth_jcp,Dornheim_PRL_2020,Dornheim_PRR_2021} and extend Eq.~(\ref{eq:Hamiltonian}) by an external static cosinusoidal perturbation of wave vector $\mathbf{q}$ and perturbation amplitude $A$,
\begin{eqnarray}\label{eq:H_perturbed}
\hat{H}_{\mathbf{q},A} = \hat{H} + 2A\sum_{l=1}^N\textnormal{cos}\left(\mathbf{q}\cdot\hat{\mathbf{r}}_l\right)\ .
\end{eqnarray}
In the limit of small perturbation amplitudes $A$, the density response is described accurately by linear-response theory, and the density profile is given by~\cite{Dornheim_PRR_2021}
\begin{eqnarray}\label{eq:n}
    n(\mathbf{r}) = n_0 + 2A\chi(\mathbf{q})\ \textnormal{cos}\left(\mathbf{q}\cdot\mathbf{r}\right)\ 
\end{eqnarray}
for uniform systems, with $\chi(\mathbf{q})$ being the linear density-response function.
Extensions of density response theory to the nonlinear regime have been discussed in detail, e.g., in Refs.~\cite{mikhailov2012,Mikhailov_Annalen,Dornheim_PRL_2020,Dornheim_PRR_2021,Dornheim_JCP_ITCF_2021,Moldabekov_JCTC_2022,Dornheim_CPP_2021,Dornheim_CPP_2022,tolias2023unravelling}, but are not covered in the present work.

As a side not, we mention that the PIMC method is perfectly capable to treat both electrons and ions on the same footing, i.e., without the Born-Oppenheimer approximation inherent to Eq.~(\ref{eq:Hamiltonian}), and without the need for an additional averaging over individual snapshots~\cite{moldabekov2023averaging} to obtain properly averaged thermodynamic properties. This, however, is not the purpose of the present work, where we intend to isolate the effects of the local ionic structure onto the electronic density, instead of averaging it out. Moreover, solving the electronic problem defined by Eqs.~(\ref{eq:Hamiltonian}) and/or Eq.~(\ref{eq:H_perturbed}) makes our PIMC results directly comparable to DFT calculations, which is important for the benchmarking of the latter. Full PIMC simulations of both electrons and ions will, therefore, be pursued in dedicated future works.

As mentioned in the introduction, it is common practice to characterize WDM in terms of the Wigner-Seitz radius $r_s=(3/4\pi n_0)^{1/3}$, with $n_0=N/\Omega$ being the mean number density. From a physical perspective, $r_s$ plays the role of a quantum coupling parameter, with $r_s\to0$ corresponding to the limit of an ideal Fermi gas and $r_s \gg 1$ indicating a strongly coupled system.
The degeneracy temperature $\Theta$ serves as an inverse degeneracy parameter, and $\Theta\ll 1$ ($\Theta\gg1$) corresponds to the fully degenerate (semi-classical~\cite{Dornheim_HEDP_2022}) limit.
A third parameter is given by the spin-polarization degree $\xi=(N^\uparrow-N^\downarrow)/N$; we limit ourselves to the unpolarized case of $\xi=0$ (i.e., $N^\uparrow=N^\downarrow$) throughout this work.

\subsection{Path integral Monte Carlo\label{sec:PIMC}}

We consider a system governed by the general Hamiltonian Eq.~(\ref{eq:H_perturbed}) in the canonical ensemble, where the number of electrons $N$, volume $\Omega$, and inverse temperature $\beta=1/k_\textnormal{B}T$ are fixed.
The canonical partition function is then readily expressed in coordinate representation as
\begin{widetext}
\begin{eqnarray}\label{eq:Z}
Z_{\beta,N,\Omega} = \frac{1}{N^\uparrow! N^\downarrow!} \sum_{\sigma^\uparrow\in S_{N^\uparrow}} \sum_{\sigma^\downarrow\in S_{N^\downarrow}} \textnormal{sgn}(\sigma^\uparrow,\sigma^\downarrow) \int d\mathbf{R} \bra{\mathbf{R}} e^{-\beta\hat H} \ket{\hat{\pi}_{\sigma^\uparrow}\hat{\pi}_{\sigma^\downarrow}\mathbf{R}}\ ,
\end{eqnarray}
\end{widetext}
where the summation over all possible permutations $\sigma^i$ from the respective permutation group $S_{N^i}$ (with $i\in\{\uparrow,\downarrow\}$) taken together with the sign function $\textnormal{sgn}(\sigma^\uparrow,\sigma^\downarrow)$ and the permutation operators $\hat{\pi}_{\sigma^\uparrow}\hat{\pi}_{\sigma^\downarrow}$ ensures the correct fermionic anti-symmetry with respect to the exchange of particle coordinates.
While Eq.~(\ref{eq:Z}) is formally exact, it cannot be evaluated in practice as the kinetic and potential contributions to the total Hamiltonian do not commute. The basic idea behind the PIMC method~\cite{cep,Berne_JCP_1982,Takahashi_Imada_PIMC_1984} is to utilize a well-known semi-group property of the density operator $\hat\rho = e^{-\beta\hat{H}}$, which makes it possible to express Eq.~(\ref{eq:Z}) as a combination of $P$ integrals in coordinate space at $P$-times the original temperature. For sufficiently large $P$, one can then introduce a suitable high-temperature factorization of $\hat{\rho}$, and the associated factorization error can be made arbitrarily small by increasing $P$. For systems where $\hat{V}$ is bounded from below, the convergence of this approach is ensured by the well-known Trotter formula~\cite{trotter,kleinert2009path}. 
This, however, is not the case for the present hydrogen problem, where the Coulomb attraction between an electron and a proton diverges for small distances $r$. This problem can be circumvented by incorporating the exact solution to the quantum two-body problem, which is typically known as \emph{pair approximation} in the literature~\cite{cep,MILITZER201688,Bohme_PRE_2023}.
It is a common practice to pre-compute the required two-body density matrix, and utilize a polynomial parametrization for the PIMC simulation itself to save compute time. Here, we follow the approach introduced in Refs.~\cite{MILITZER201688,Bohme_PRE_2023}.
In addition, we have also implemented the diagonal Kelbg potential~\cite{Filinov_PRE_2004}, which is based on a perturbation expansion around the exact pair density matrix. 
Both Kelbg and the full \emph{pair approximation} become exact in the limit of large $P$ (see Sec.~\ref{sec:convergence} for numerical results), although the latter converges substantially faster in practice.

To evaluate the resulting $3PN$-dimensional integral, one typically employs some implementation of the celebrated Metropolis algorithm~\cite{metropolis}. In this work, we use the extended ensemble approach introduced in Ref.~\cite{Dornheim_PRB_nk_2021}, which is a canonical adaption of the worm algorithm by Boninsegni \emph{et al.}~\cite{boninsegni1,boninsegni2}.

An additional obstacle regarding the PIMC simulation of quantum degenerate fermions (such as the electrons in warm dense hydrogen) is given by the sign function in Eq.~(\ref{eq:Z}); it leads to contributions with alternating signs, which might cancel to a large degree. This is the root cause of the notorious fermion sign problem~\cite{troyer,dornheim_sign_problem}, which leads to an exponential increase in the required compute time with increasing the system size $N$ or decreasing the temperature $T$. While the alleviation of this exponential bottleneck constitutes a highly active topic of research~\cite{Schoof_PRL_2015,Dornheim_NJP_2015,Malone_JCP_2015,Dornheim_JCP_2015,Schoof_CPP_2015,Filinov_PRE_2015,Yilmaz_JCP_2020,Hirshberg_JCP_2020,Dornheim_Bogoliubov_2020,Yilmaz_JCP_2020,Filinov_CPP_2021,Joonho_JCP_2021,Xiong_JCP_2022}, no general solution appears to be realistic at the present time. In this work, we carry out direct PIMC simulations that are subject to the full sign problem and, therefore, exact within the given Monte Carlo error bars.

An alternative strategy has been introduced by Ceperley~\cite{Ceperley1991} in the form of the fixed-node approximation. On the one hand, this approach allows one to formally avoid the sign problem, and, therefore, to perform simulations over substantial parts of the WDM regime. This has allowed Militzer and co-workers to present such restricted PIMC results for a variety of different elements at WDM conditions~\cite{militzer1,Driver_PRL_2012,Driver_PRB_2016}, which have subsequently been compiled into an extensive equation-of-state database~\cite{Militzer_PRE_2021}.
On the other hand, the fixed-node approximation is uncontrolled in practice, and its accuracy generally remains unclear. Indeed, Schoof \emph{et al.}~\cite{Schoof_PRL_2015} have found that the exchange--correlation energy can exhibit errors exceeding $10\%$ at high densities, which has subsequently been corroborated by an independent group~\cite{Joonho_JCP_2021}. In contrast, Dornheim \emph{et al.}~\cite{Dornheim_PRB_nk_2021,Dornheim_PRE_2021} have found good agreement for the momentum distribution at moderate temperatures.
The quasi-exact PIMC results for warm dense hydrogen that are presented here thus open up the intriguing possibility to rigorously assess the accuracy of the fixed-node approximation (as well as any other simulation method, including DFT) for a real system, i.e., beyond the UEG, on the electronic nanoscale.

\begin{figure}\centering
\includegraphics[width=0.45\textwidth]{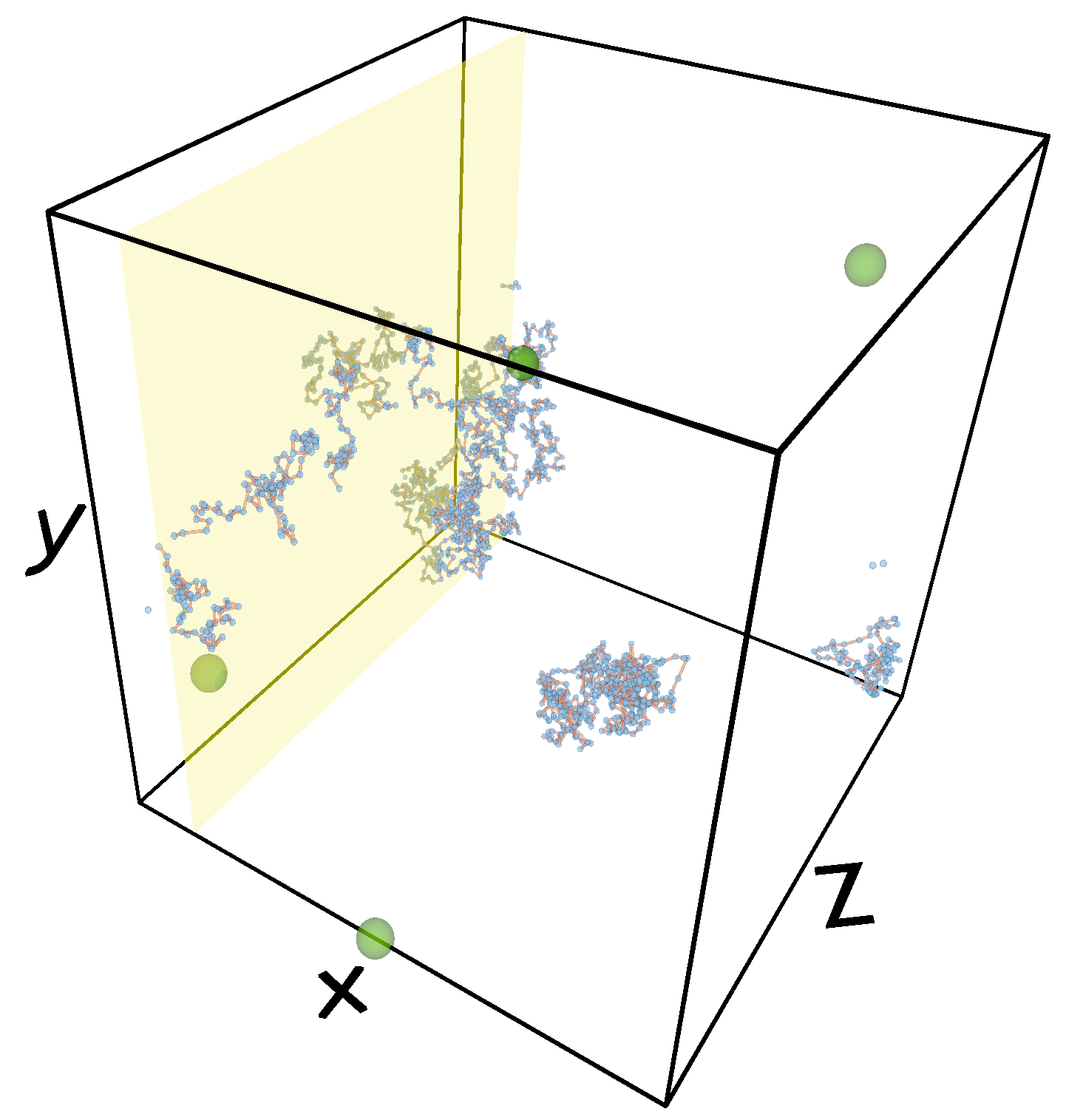}
\caption{\label{fig:Snap_N4_rs4_theta1}
Snapshot of a PIMC simulation with $N=4$, $r_s=4$, $\Theta=1$ (i.e., $T=3.13\,$eV), and $P=500$. The green orbs depict the protons, and the red-blue paths visualize a given electronic configuration. The yellow surface in the $y$-$z$-plane is investigated in more detail in Fig.~\ref{fig:Layer2_PA1000_N4_rs4_theta1}.
}
\end{figure}

\section{Results\label{sec:results}}

\subsection{Convergence\label{sec:convergence}}

Let us begin our investigation with an in-depth analysis of the convergence of the PIMC results with the number of high-temperature factors $P$. To this end, we carry out simulations with $N=4$ at $r_s=4$ and $\Theta=1$. This corresponds to dilute, strongly coupled hydrogen which can be realized in experiments with hydrogen jets~\cite{Zastrau}. The expected high degree of localization of the electrons around the protons~\cite{Bohme_PRE_2023} makes this density particularly challenging with respect to the convergence with $P$.
In Fig.~\ref{fig:Snap_N4_rs4_theta1}, we show a snapshot from a corresponding PIMC simulation with $P=500$ high-temperature factors, with the green orbs representing the protons and the red-blue paths depicting a particular electron configuration.

\begin{figure}\centering
\includegraphics[width=0.475\textwidth]{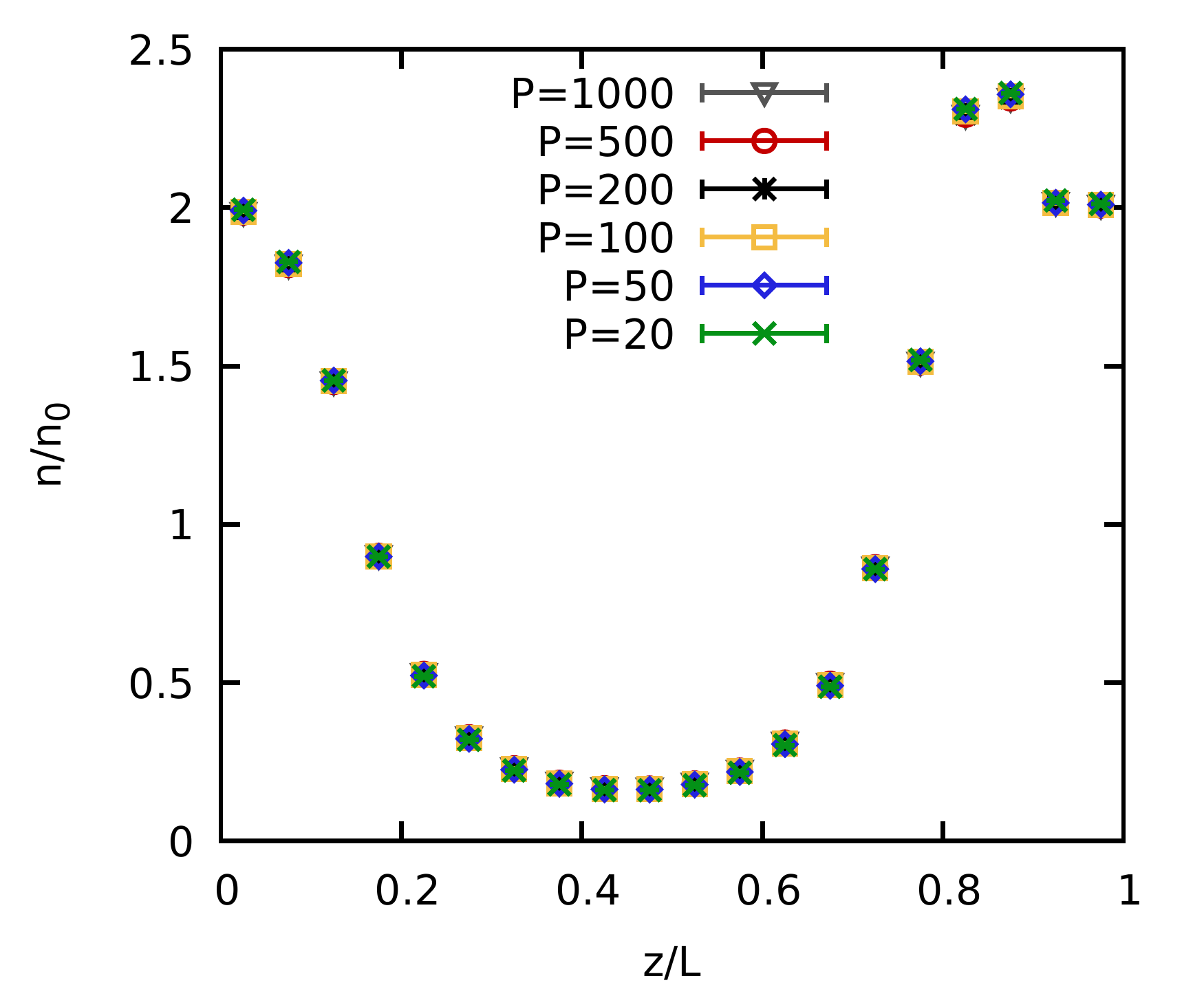}
\caption{\label{fig:density_strip_N4_rs4_theta1}
Density strip along the $z$-direction for $r_s=4$ and $\Theta=1$, computed from PIMC simulations using the pair approximation for different numbers of high-temperature factors $P$.
}
\end{figure}

As a first example, we investigate the density along the $z$-direction (i.e., averaged over $x$ and $y$) in Fig.~\ref{fig:density_strip_N4_rs4_theta1}. Specifically, the different symbols show results from individual PIMC simulations using the pair approximation for different numbers of high-temperature factors $P$. First, we find a high degree of localization as the electronic density nearly vanishes in between the protons. Second, hardly any factorization error is visible on the depicted scale.

\begin{figure*}\centering
\includegraphics[width=0.475\textwidth]{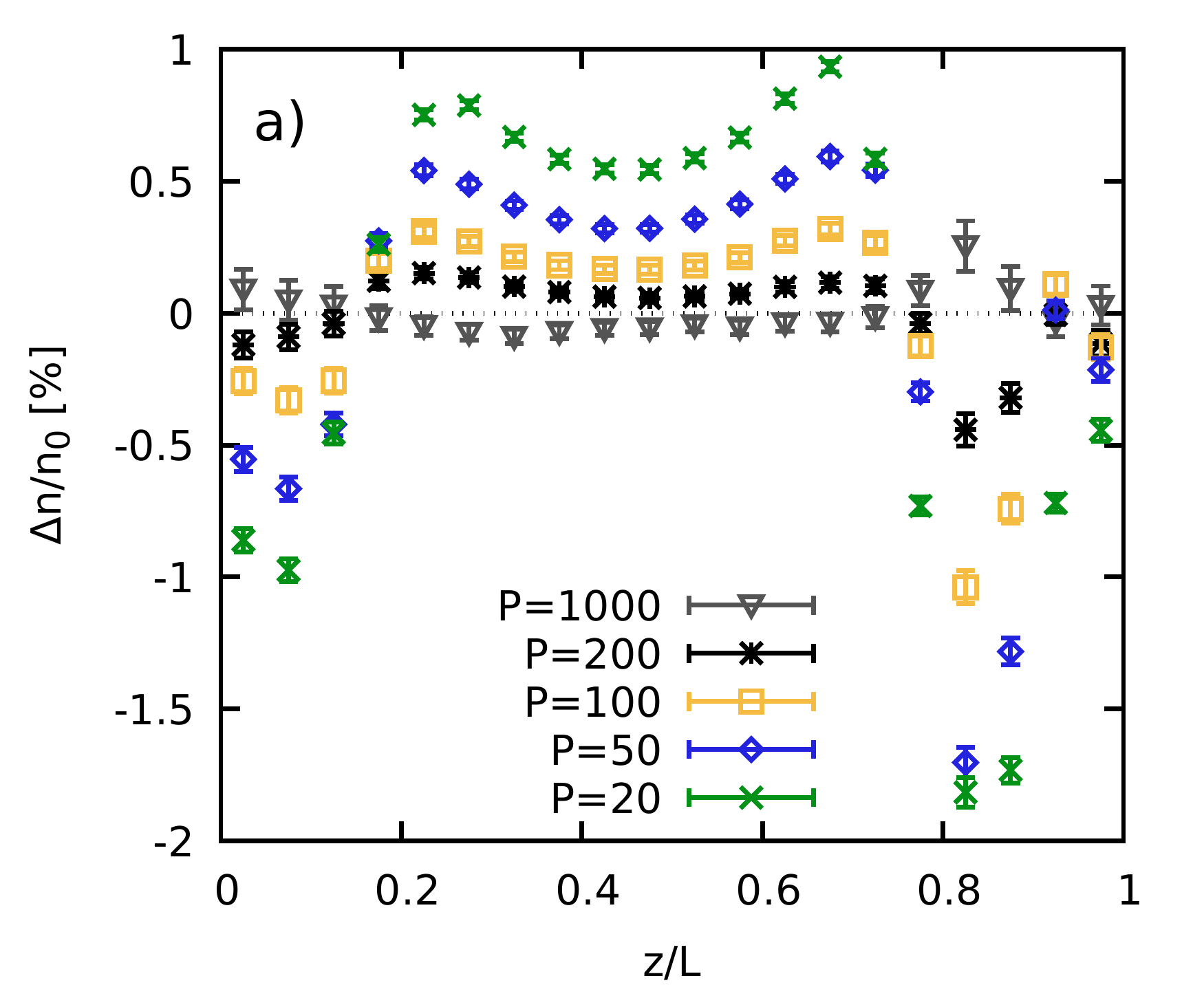}\includegraphics[width=0.475\textwidth]{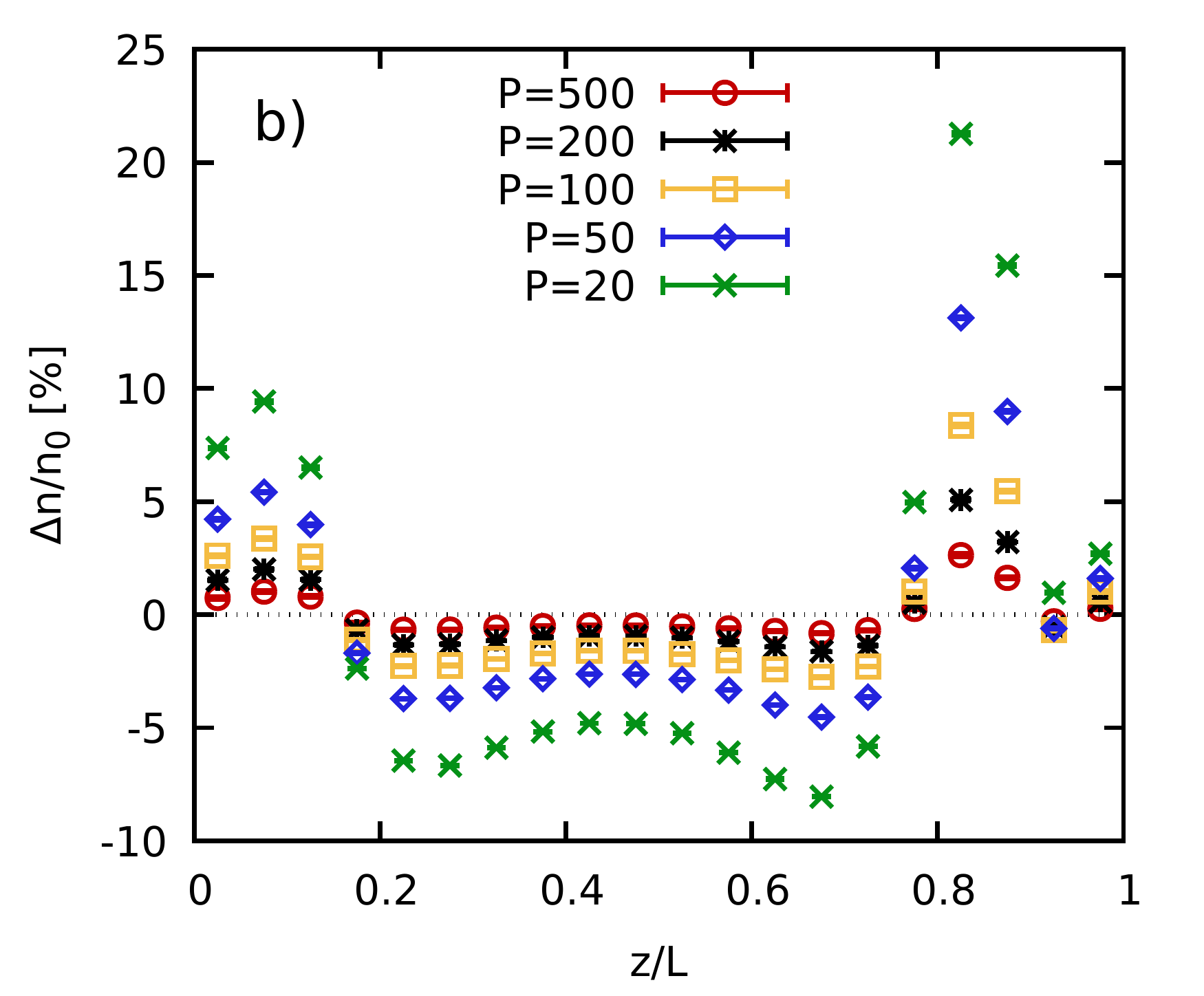}
\caption{\label{fig:convergence_density_strip_N4_rs4_theta1}
Relative difference in the density along the $z$-direction towards $P=500$ pair-approximation reference data. a) pair approximation; b) diagonal Kelbg potential.
}
\end{figure*}

To get more quantitative insights into the convergence behaviour, we analyze the difference towards a reference data set that has been computed with the pair approximation for $P=500$ in Fig.~\ref{fig:convergence_density_strip_N4_rs4_theta1}. More specifically, panel a) shows the relative deviation for different $P$, and we observe a monotonous convergence with increasing $P$. Remarkably, we find a small factorization error that is bounded by $\pm2\%$ of the average density $n_0$ even for as few as $P=20$ high temperature factors; the results for $P=1000$ and $P=500$ cannot be distinguished within the given level of statistical uncertainty. In other words, PIMC simulations with $P=500$ pair-approximation propagators, which are used in Secs.~\ref{sec:low_density} and \ref{sec:density} below, give us an accuracy of $\sim0.1\%$ in the density.

In Fig.~\ref{fig:convergence_density_strip_N4_rs4_theta1}b), we repeat this analysis, but consider the difference to the $P=500$ pair-approximation reference data set to PIMC calculations using the diagonal Kelbg potential. While we again find a monotonously decreasing factorization error in these simulations, the factorization error of the Kelbg approximation is an order of magnitude larger compared to panel a). This clearly demonstrates the superior performance of the full pair approximation and is consistent to previous investigations in Refs.~\cite{Bohme_PRE_2023,Filinov_PRE_2004}.

\begin{figure}\centering
\hspace*{-0.5cm}\includegraphics[width=0.55\textwidth]{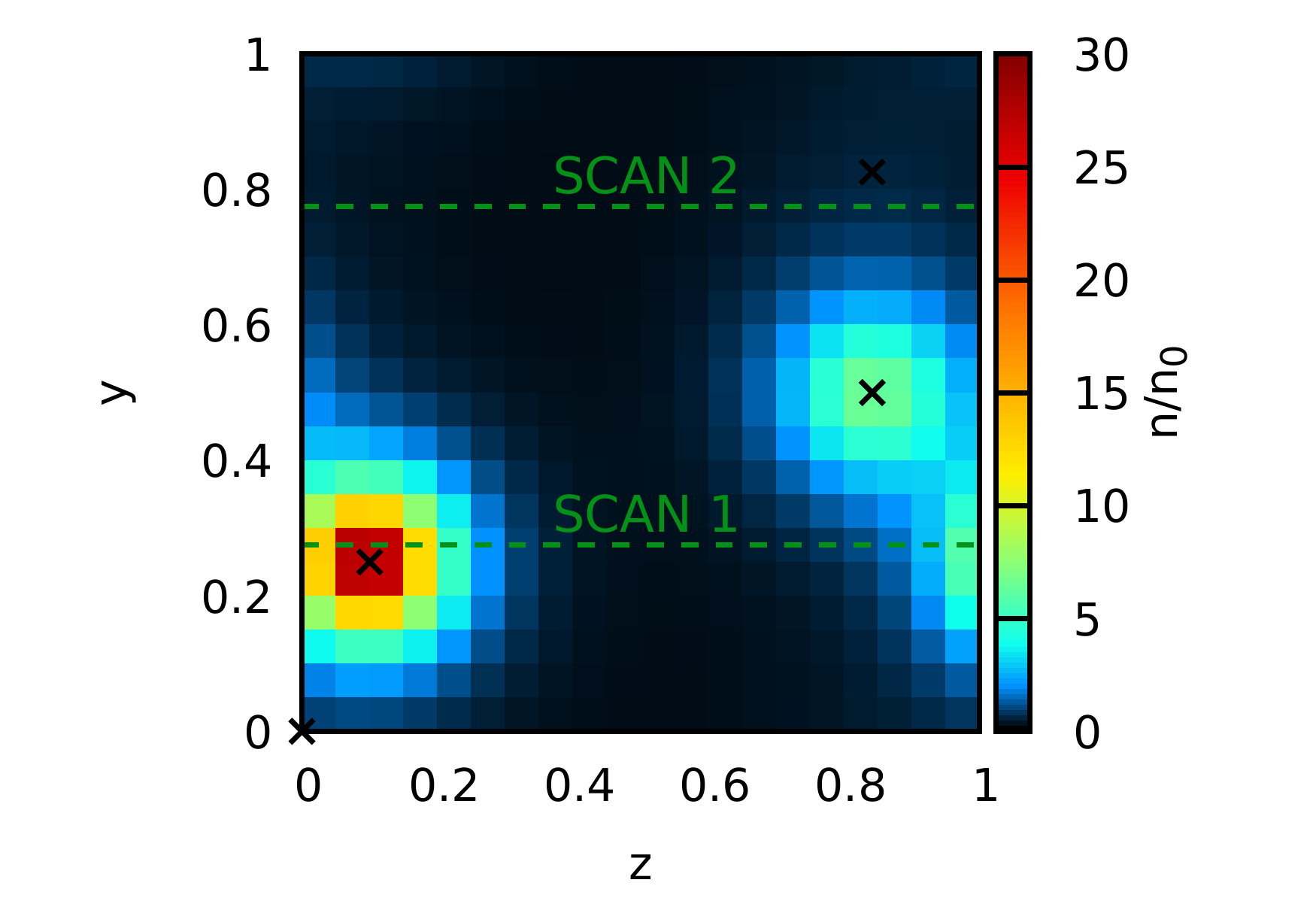}
\caption{\label{fig:Layer2_PA1000_N4_rs4_theta1}
Top: electronic density in the $y$-$z$-plane for $P=1000$, see the yellow surface in Fig.~\ref{fig:Snap_N4_rs4_theta1}. The dashed green lines depict to scan lines that are investigated in Fig.~\ref{fig:SCAN_Layer2_PA1000_N4_rs4_theta1}.
}
\end{figure}

While the observed high fidelity of the pair approximation is promising, potential factorization errors might still have been masked by the averaging over the $x$- and $y$-directions for the density strips shown in Figs.~\ref{fig:density_strip_N4_rs4_theta1} and~\ref{fig:convergence_density_strip_N4_rs4_theta1}. Indeed, one would expect any factorization errors to be particularly manifest in the vicinity of the protons, where the gradient in the electron density is large. At the same time, we stress that accurate PIMC data are also particularly important precisely in this region to benchmark other methods, most notably density functional theory.
To rigorously assess the convergence of our PIMC simulations with $P$, we consider the density in the $z$-$y$-plane for a value of $x$ that is identical to the location of a proton, see the yellow surface in Fig.~\ref{fig:Snap_N4_rs4_theta1}.

Let us first consider the PIMC results for the density itself, which is shown in Fig.~\ref{fig:Layer2_PA1000_N4_rs4_theta1}. This nicely illustrates the high degree of localization around the in-plane proton (bottom left corner), which had been mostly averaged out for the density strip shown in Fig.~\ref{fig:density_strip_N4_rs4_theta1}. Indeed, the density in the direct vicinity of the proton is increased by a factor of almost $30$ compared to the average density $n_0$. Consequently, the density nearly vanishes between the protons.

\begin{figure}\centering\hspace*{-0.5cm}\includegraphics[width=0.55\textwidth]{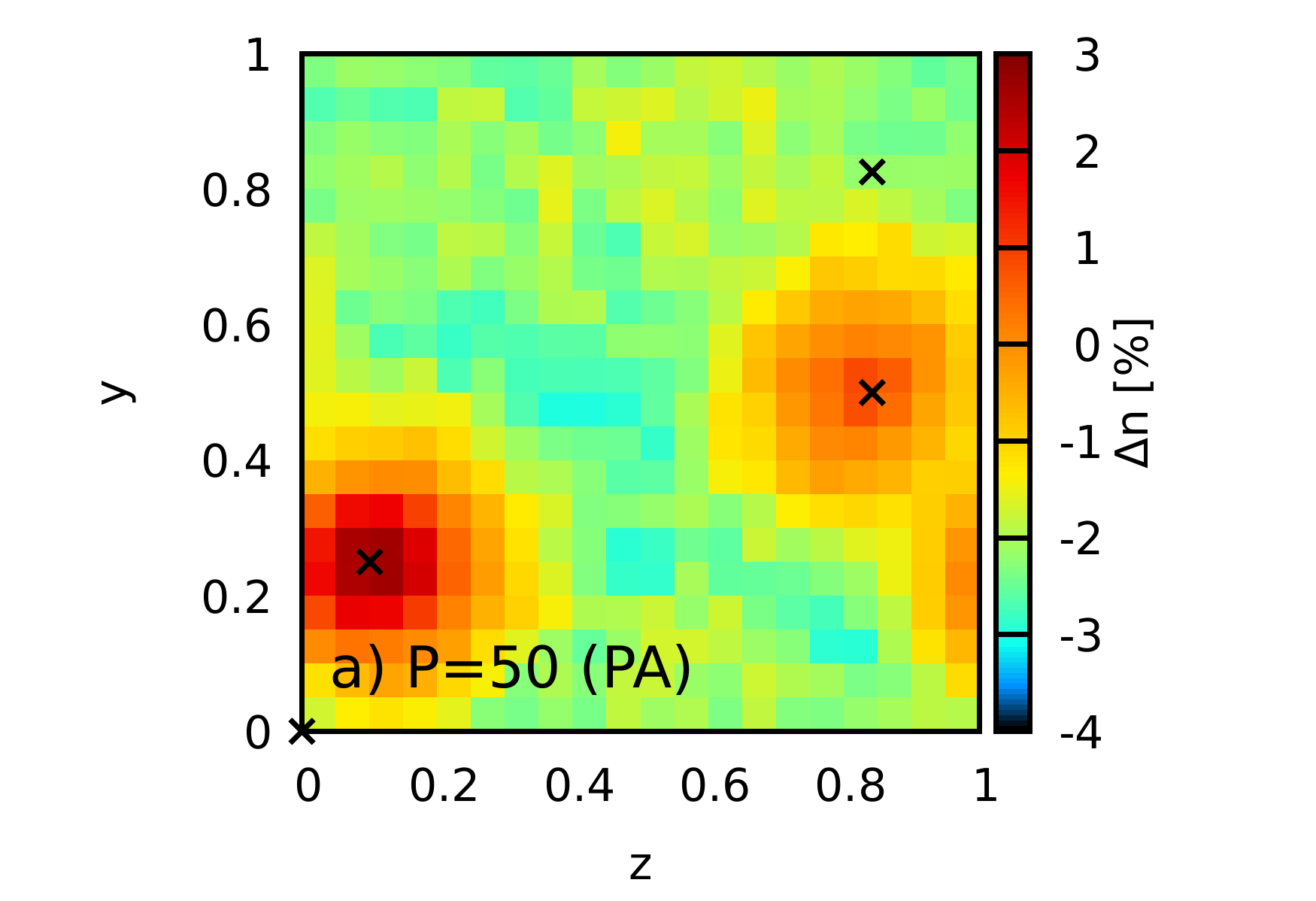}\\\vspace*{-0.75cm}\hspace*{-0.5cm}\includegraphics[width=0.55\textwidth]{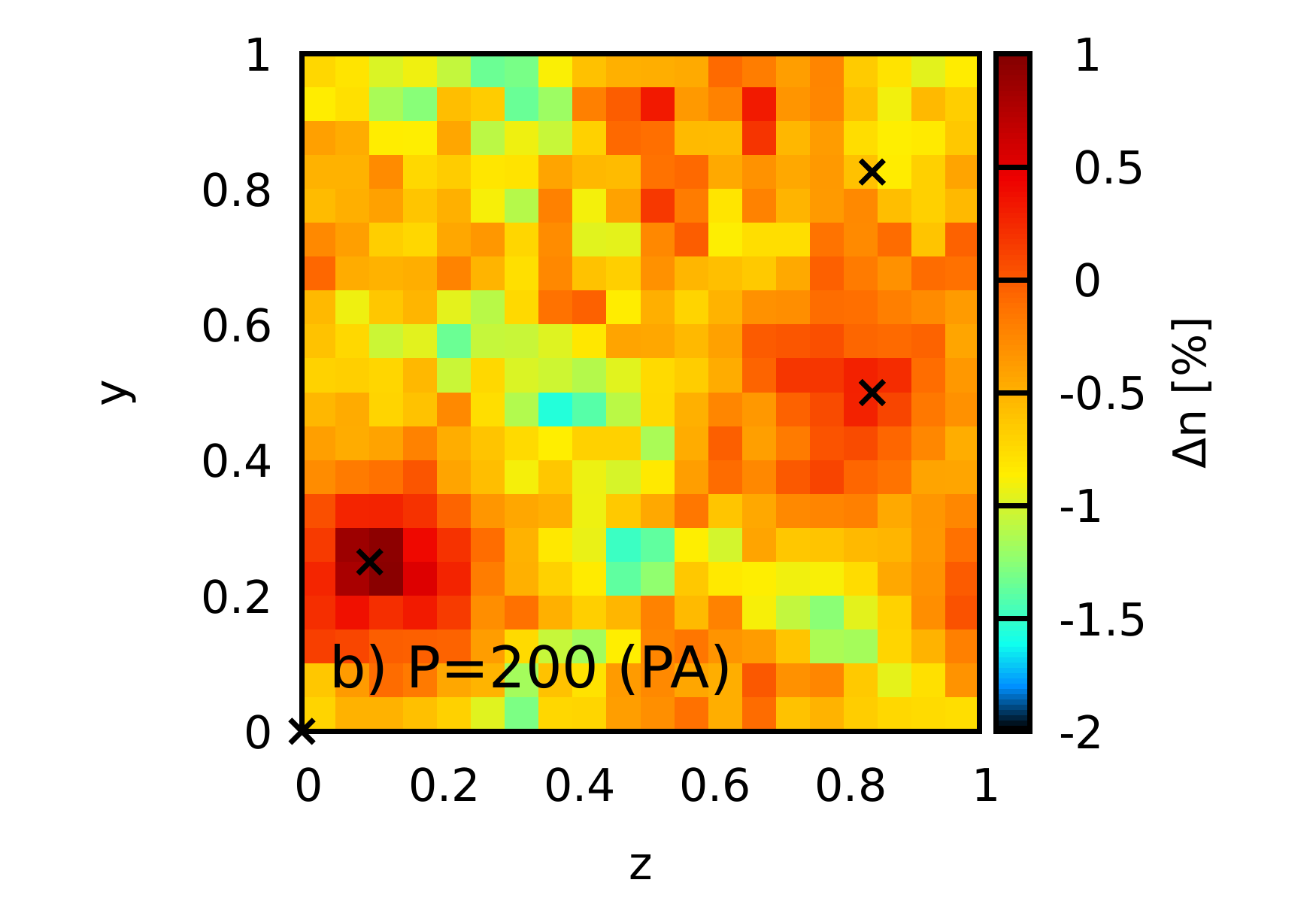}
\caption{\label{fig:Delta_Layer2_PA1000_N4_rs4_theta1}
Relative difference (in $\%$) towards PIMC reference data with $P=1000$ (cf.~Fig.~\ref{fig:Layer2_PA1000_N4_rs4_theta1}) for a) $P=50$ and b) $P=200$ pair-approximation factors.
}
\end{figure}

\begin{figure*}\centering
\includegraphics[width=0.5\textwidth]{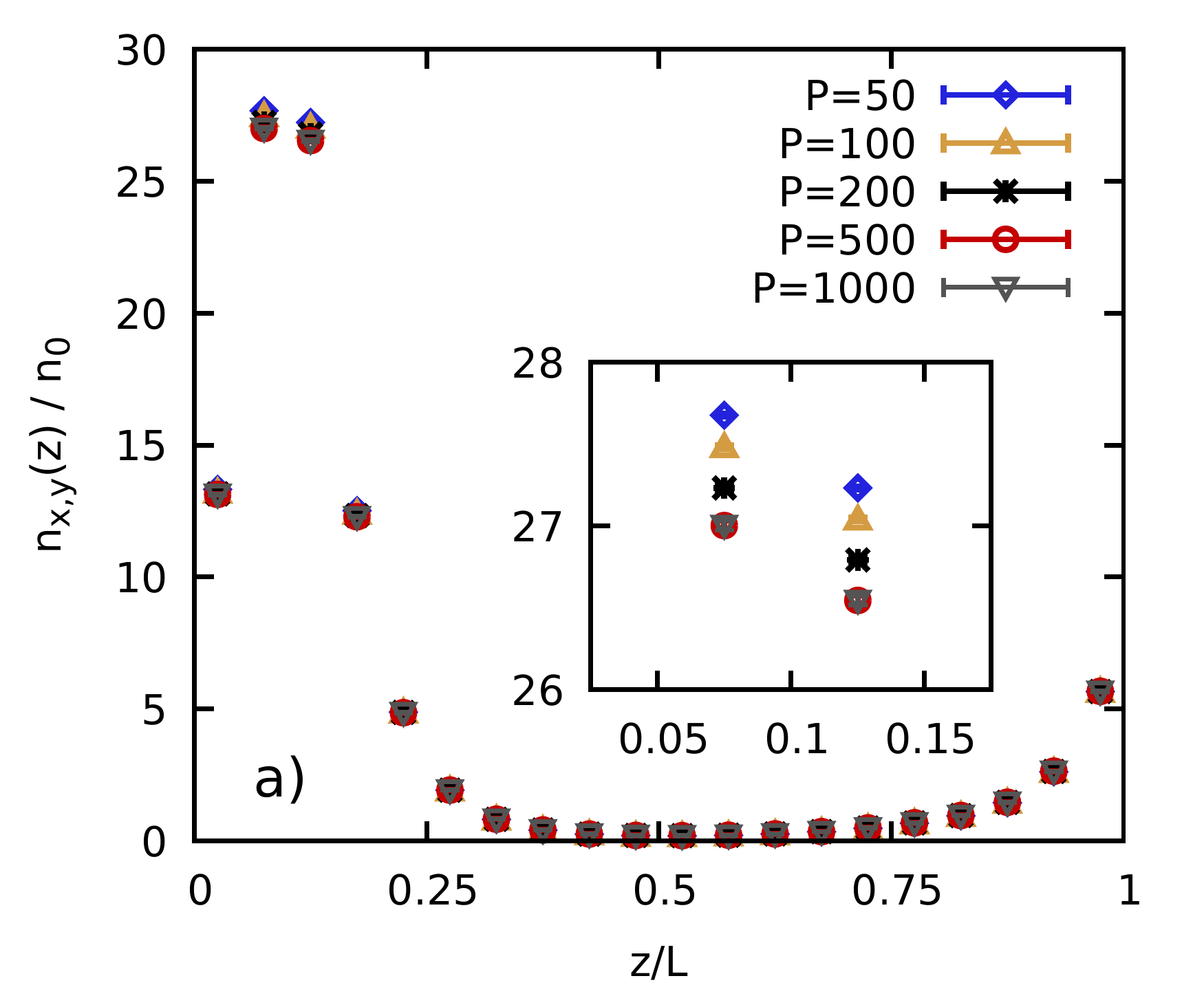}\includegraphics[width=0.5\textwidth]{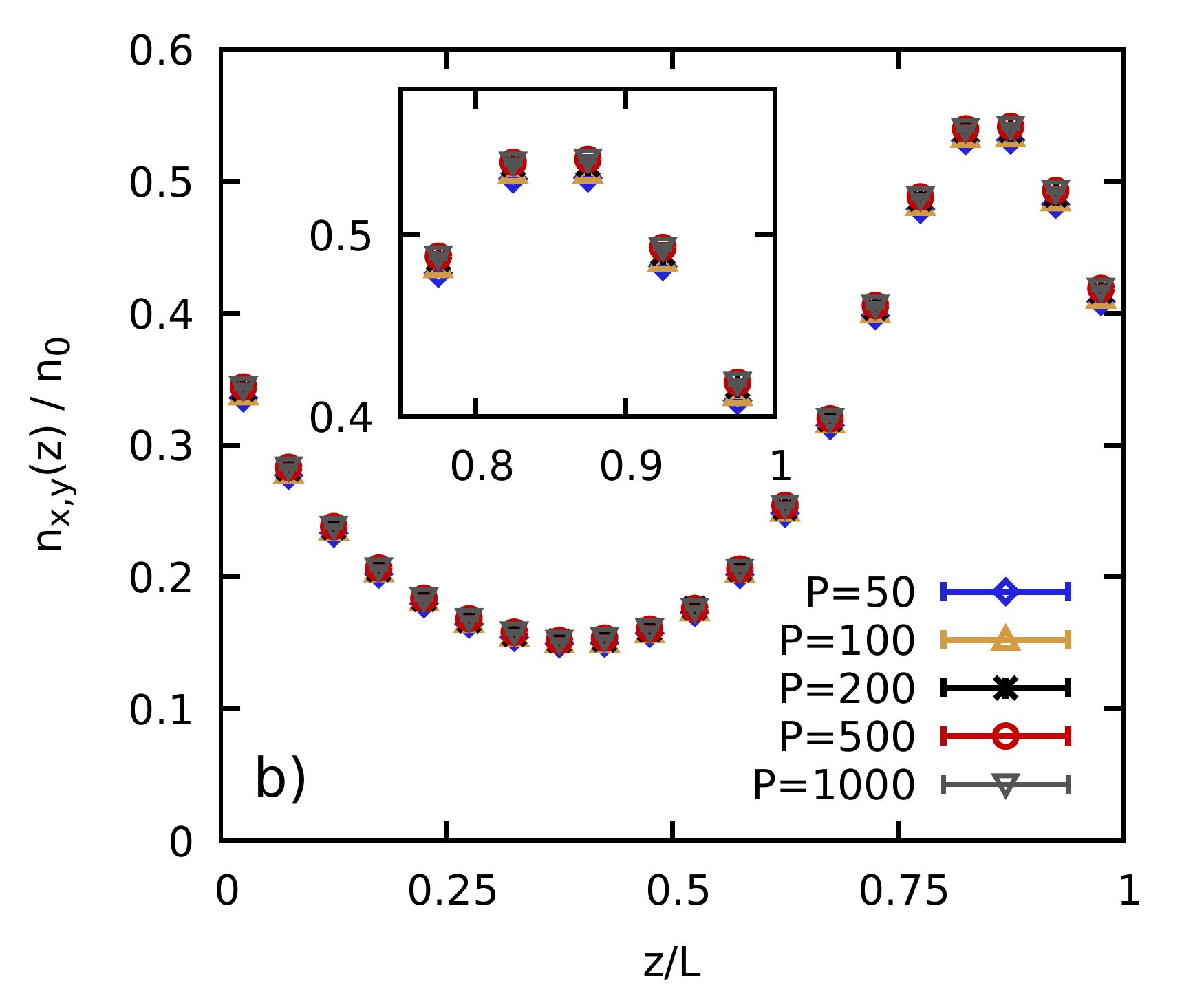}
\caption{\label{fig:SCAN_Layer2_PA1000_N4_rs4_theta1}
Scan lines over the density in the $y$-$z$-plane shown in Fig.~\ref{fig:Layer2_PA1000_N4_rs4_theta1}, computed for different numbers of high-temperature factors $P$ using the pair approximation.
}
\end{figure*}

To investigate the spatially resolved manifestation of the factorization error in our PIMC simulations, we show the relative difference towards reference data with $P=1000$ for $P=50$ (top) and $P=200$ (bottom). From the top panel, we see that the relative propagator error has a similar magnitude around the protons and in between, although with a different sign. Specifically, the localization around the protons is overestimated by about $3\%$ for $P=50$. Increasing the number of high-temperature factors to $P=200$ leads to a maximum error of $1\%$ around the proton in the bottom left corner. Hardly any propagator error can be resolved within the given error bars for $P=500$, which further substantiates our previous estimate regarding the capability of PIMC to provide the electronic density with an accuracy of $\sim0.1\%$ over the entire system.

This becomes even more clear in Fig.~\ref{fig:SCAN_Layer2_PA1000_N4_rs4_theta1}, where we show the density along two scan lines (see the dashed green lines in Fig.~\ref{fig:Layer2_PA1000_N4_rs4_theta1}) for different values of $P$. Panel a) includes the direct vicinity of a proton, where the factorization error is most pronounced in absolute terms (although not in relative terms, cf.~Fig.~\ref{fig:Delta_Layer2_PA1000_N4_rs4_theta1}). This region is magnified in the inset, thereby giving us additional insights into the convergence with $P$. No difference between $P=500$ (red circles) and $P=1000$ (grey triangles) can be resolved within the Monte Carlo error bars. In panel b), we show the same analysis for the second scan line over a region without a proton and, therefore, with low electronic density. In this region, the density gradients are small, and hardly any factorization error can be resolved even for $P=50$.

\begin{figure}\centering
\includegraphics[width=0.5\textwidth]{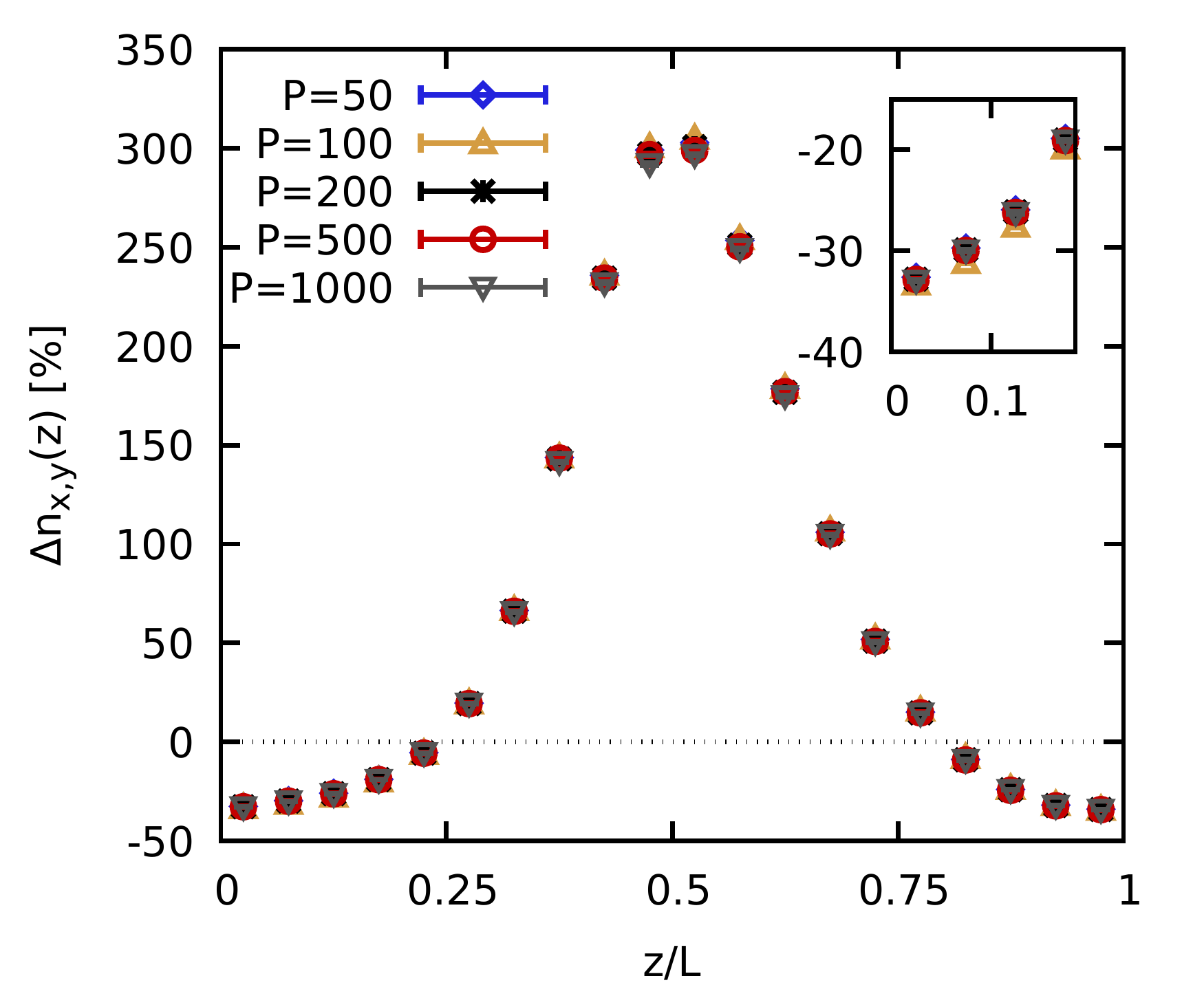}
\caption{\label{fig:Diff_P500_qz1_A0p1_Layer_2}
Induced relative density change along scan line 1 (cf.~Fig.~\ref{fig:Layer2_PA1000_N4_rs4_theta1}) due to an external harmonic perturbation with $\mathbf{q}=2\pi/L(0,0,1)^T$ and $A=0.1$, see Eq.~(\ref{eq:H_perturbed}), for different numbers of high-temperature factors $P$.
}
\end{figure}

Let us conclude this convergence study by considering the static electronic density response to an external cosinusoidal perturbation. In Fig.~\ref{fig:Diff_P500_qz1_A0p1_Layer_2}, we show the relative change in the density along scan line 1 (cf.~Fig.~\ref{fig:Layer2_PA1000_N4_rs4_theta1}) between the unperturbed system, and a harmonically perturbed snapshot calculation with $A=0.1$ and $\mathbf{q}=2\pi/L(0,0,1)^T$. Let us postpone the physical interpretation, and exclusively focus on the convergence with $P$. If anything, we find that factorization errors seem to cancel to a large degree between the perturbed and unperturbed results for the density, and hardly any differences can be resolved even for as few as $P=50$.

\subsection{Density response of low-density hydrogen\label{sec:low_density}}

\begin{figure}\centering
\includegraphics[width=0.45\textwidth]{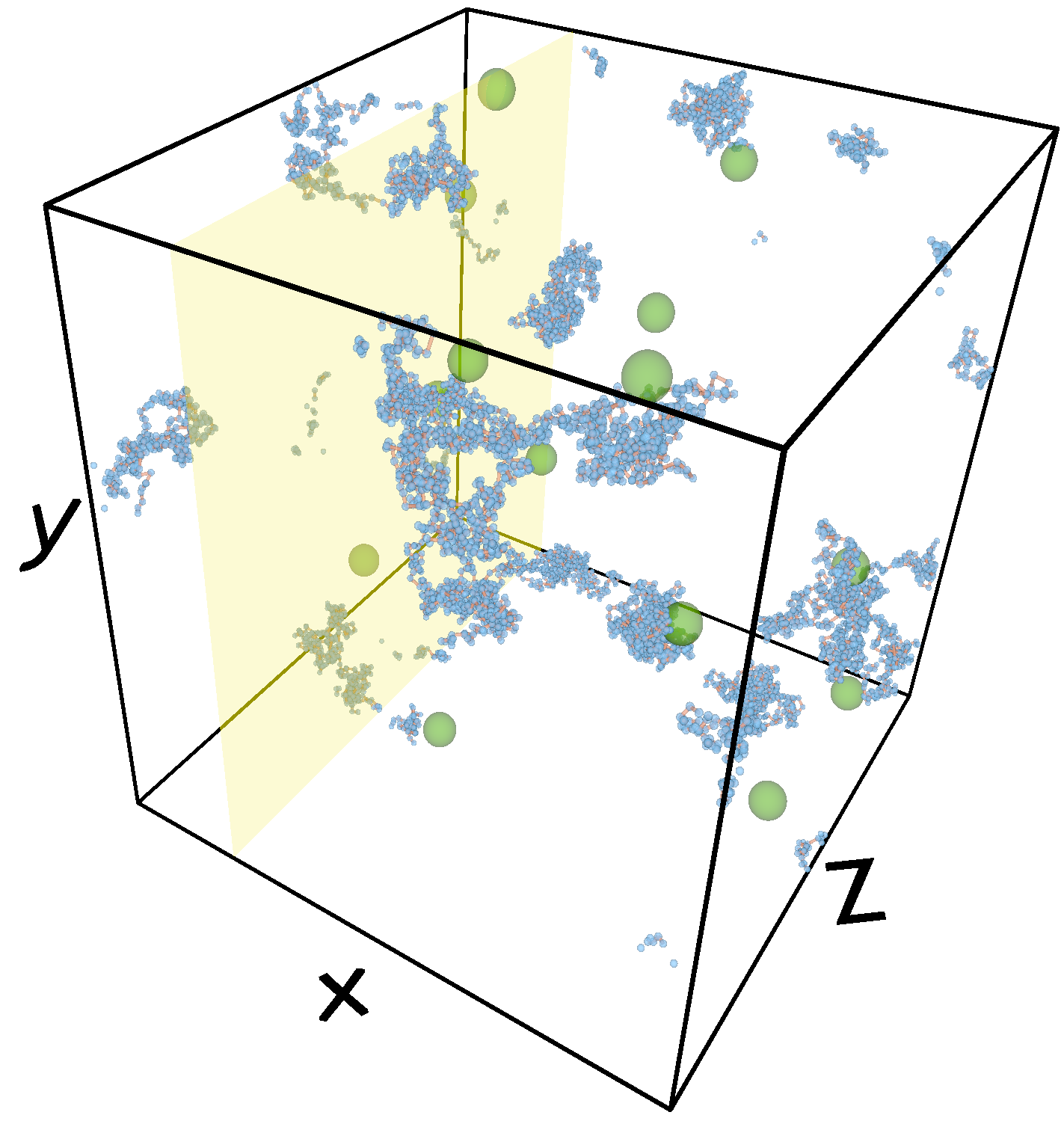}
\caption{\label{fig:Snap_N14_rs4_theta1}
Snapshot from a PIMC simulation with $N=14$, $r_s=4$, $\Theta=1$, and $P=500$. The green orbs depict the protons, and the red-blue paths visualize a given electronic configuration. The yellow surface in the $y$-$z$-plane is investigated in more detail in Figs.~\ref{fig:N14_rs4_theta1_unperturbed}, \ref{fig:panel} and \ref{fig:C2P_rs}.
}
\end{figure}

In the following, we will investigate in more detail the physical impact of the ions on the electronic density response of hydrogen on the nanoscale. We again start by considering the comparably dilute regime with $r_s=4$, where the impact of the ions is most pronounced. From a physical perspective, these conditions might give rise to interesting phenomena such as a recently predicted roton-type feature in the dynamic structure factor~\cite{hamann2023prediction}, see also Refs.~\cite{dornheim_dynamic,Dornheim_Nature_2022,Dornheim_Force_2022,Takada_PRB_2016,koskelo2023shortrange} for studies of this effect in the UEG.
In addition, this regime constitutes a challenging benchmark for other simulation methods such as DFT due to the large impact of electronic exchange--correlation effects~\cite{low_density1,low_density2}; this is a direct consequence of the role of the Wigner-Seitz radius as the quantum coupling parameter.
A corresponding snapshot from a PIMC simulation with $N=14$, $\Theta=1$, and $P=500$ is shown in Fig.~\ref{fig:Snap_N14_rs4_theta1}.

\begin{figure}\centering
\hspace*{-0.5cm}\includegraphics[width=0.55\textwidth]{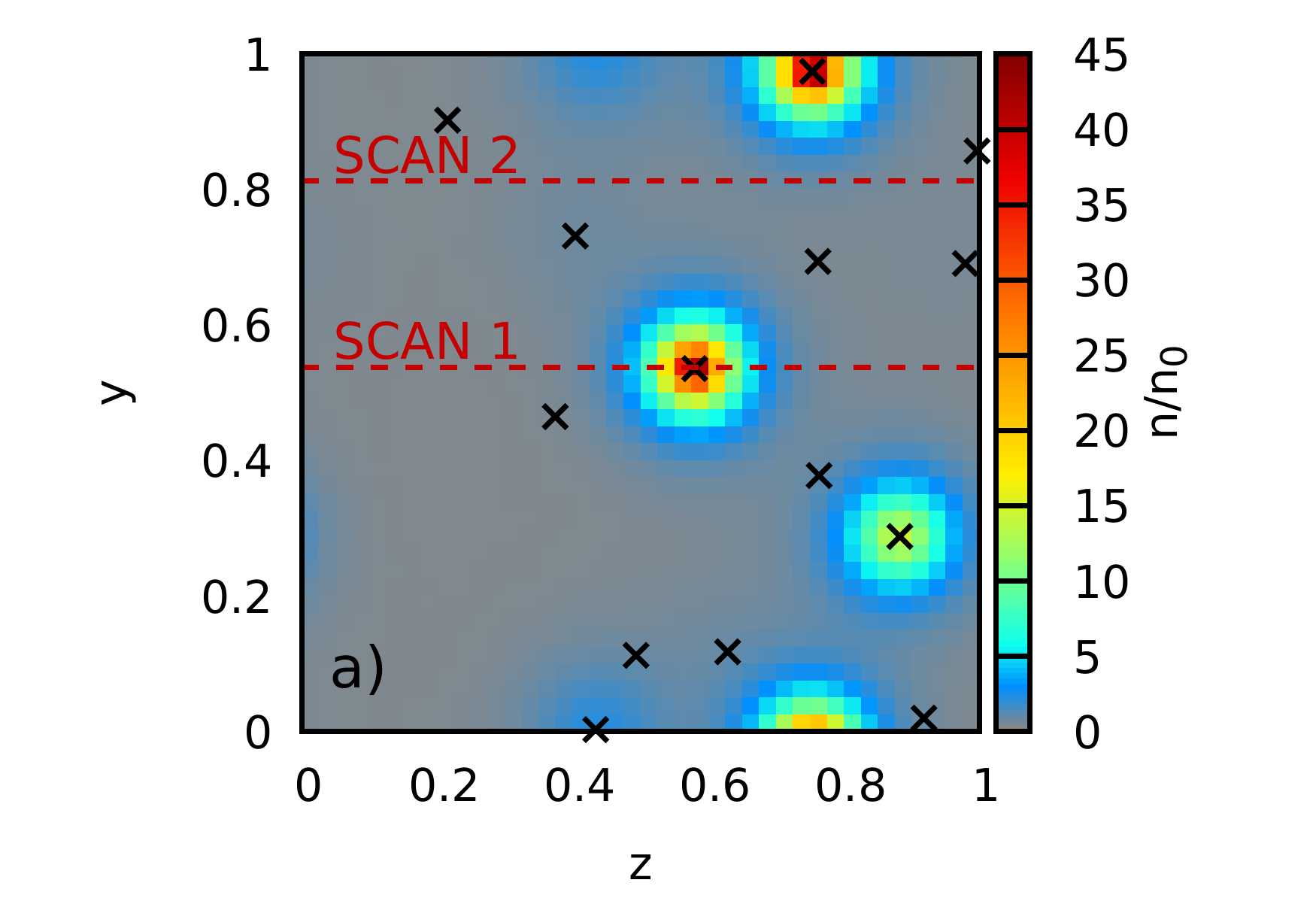}\\\vspace*{-0.85cm}\hspace*{-0.5cm}\includegraphics[width=0.55\textwidth]{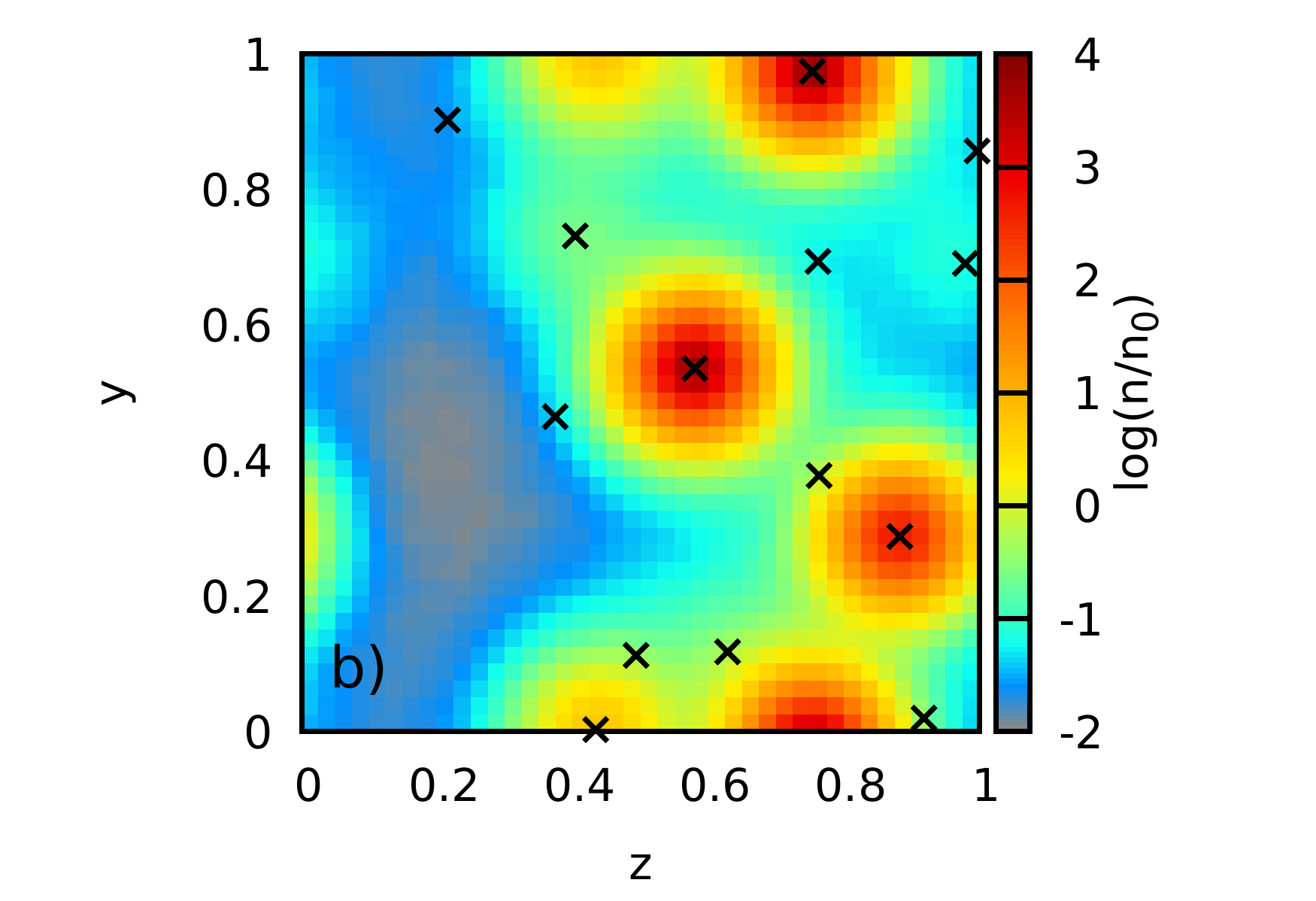}
\caption{\label{fig:N14_rs4_theta1_unperturbed}
PIMC results for the (unperturbed, i.e., $A=0$) electronic density in the $z$-$y$-plane (cf.~the yellow surface in Fig.~\ref{fig:Snap_N14_rs4_theta1}) for $N=14$, $r_s=4$, $\Theta=1$, and $P=500$. a) density; b) logarithm of the density. 
}
\end{figure}

In the top panel of Fig.~\ref{fig:N14_rs4_theta1_unperturbed}, we show PIMC results for the electronic density in the $z$-$y$-plane (cf.~the yellow surface in Fig.~\ref{fig:Snap_N14_rs4_theta1}). We find a high degree of localization around the two in-plane protons, with a relative increase in the density compared to the average value of $n_0$ of around $40$. In addition, there appear regions with an approximately vanishing density in between.
To get a better insight into the latter, we also show the logarithm of the density in the bottom panel of Fig.~\ref{fig:N14_rs4_theta1_unperturbed}, which reveals a richer structure.


\begin{figure*}\centering
\hspace*{-0.5cm}\includegraphics[width=0.55\textwidth]{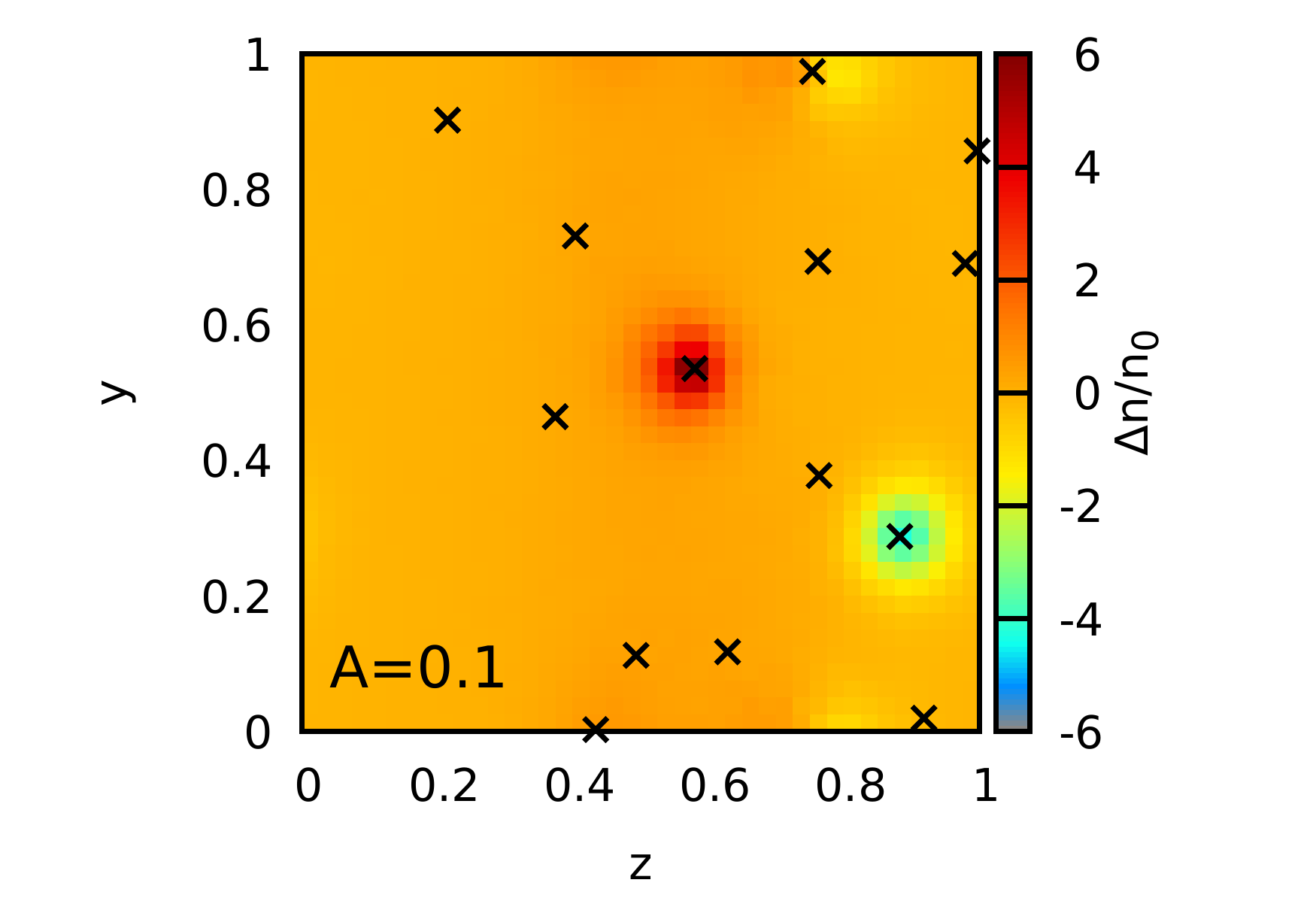}\hspace*{-1cm}\includegraphics[width=0.55\textwidth]{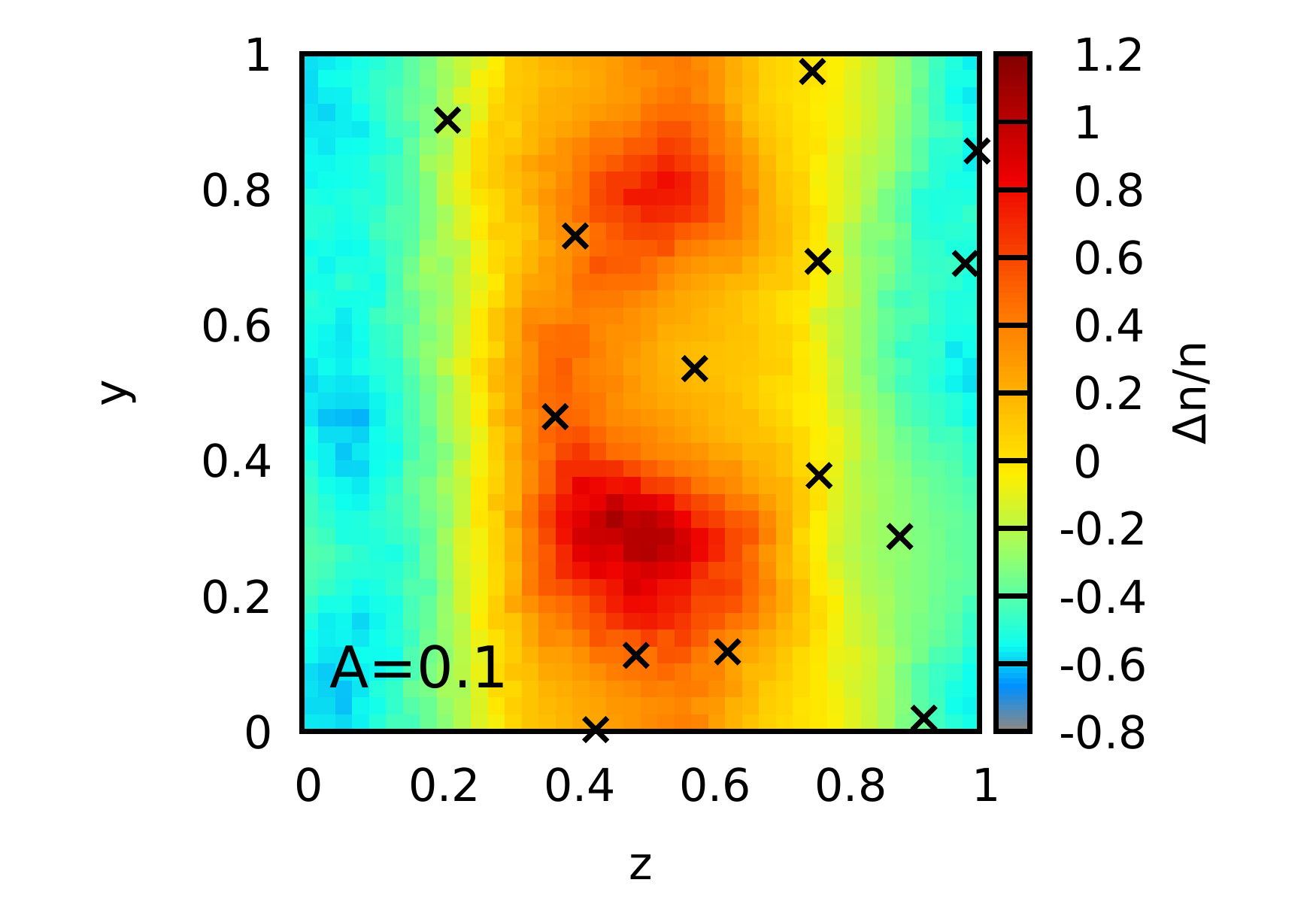}
\hspace*{-0.5cm}\includegraphics[width=0.55\textwidth]{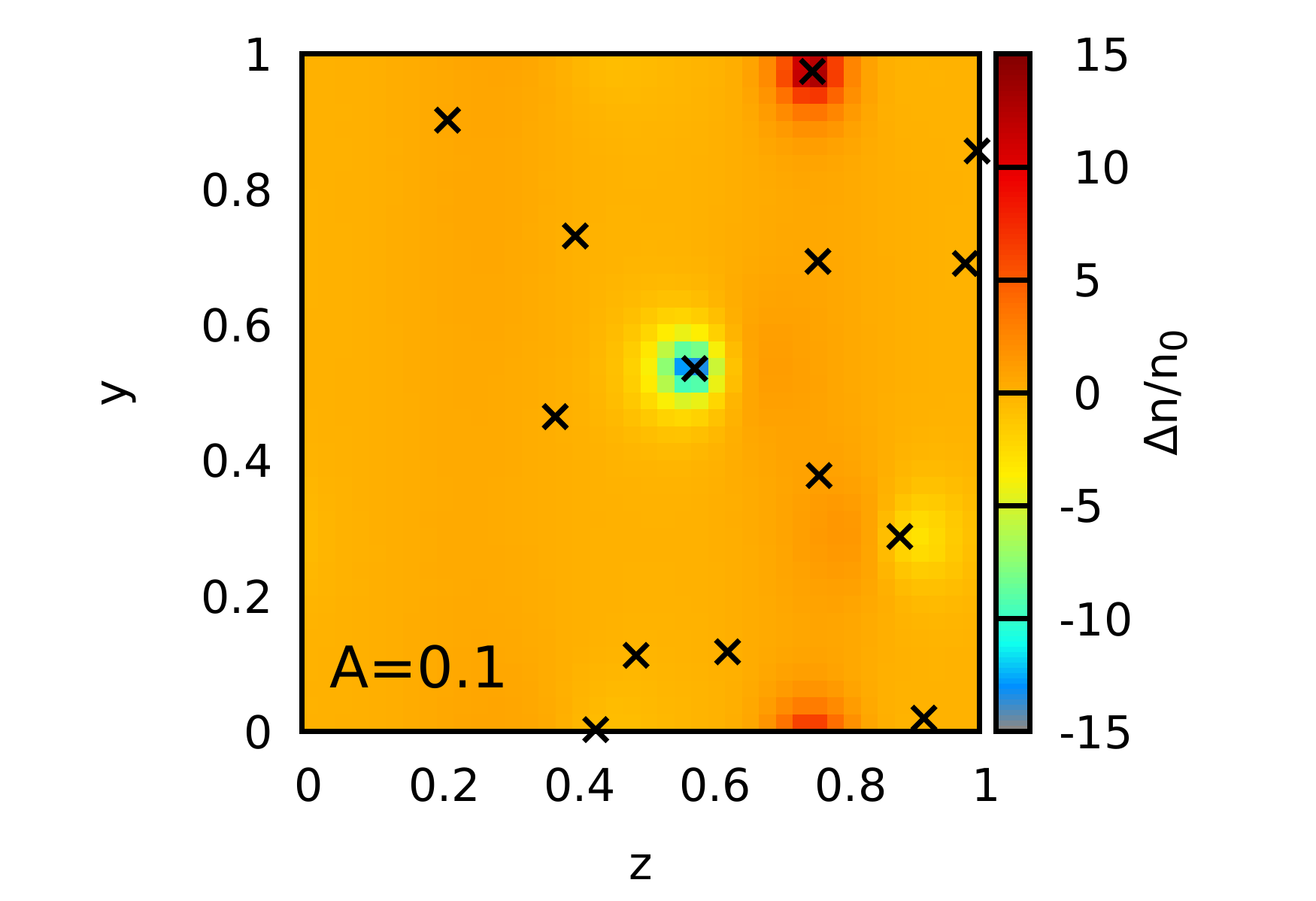}\hspace*{-1cm}\includegraphics[width=0.55\textwidth]{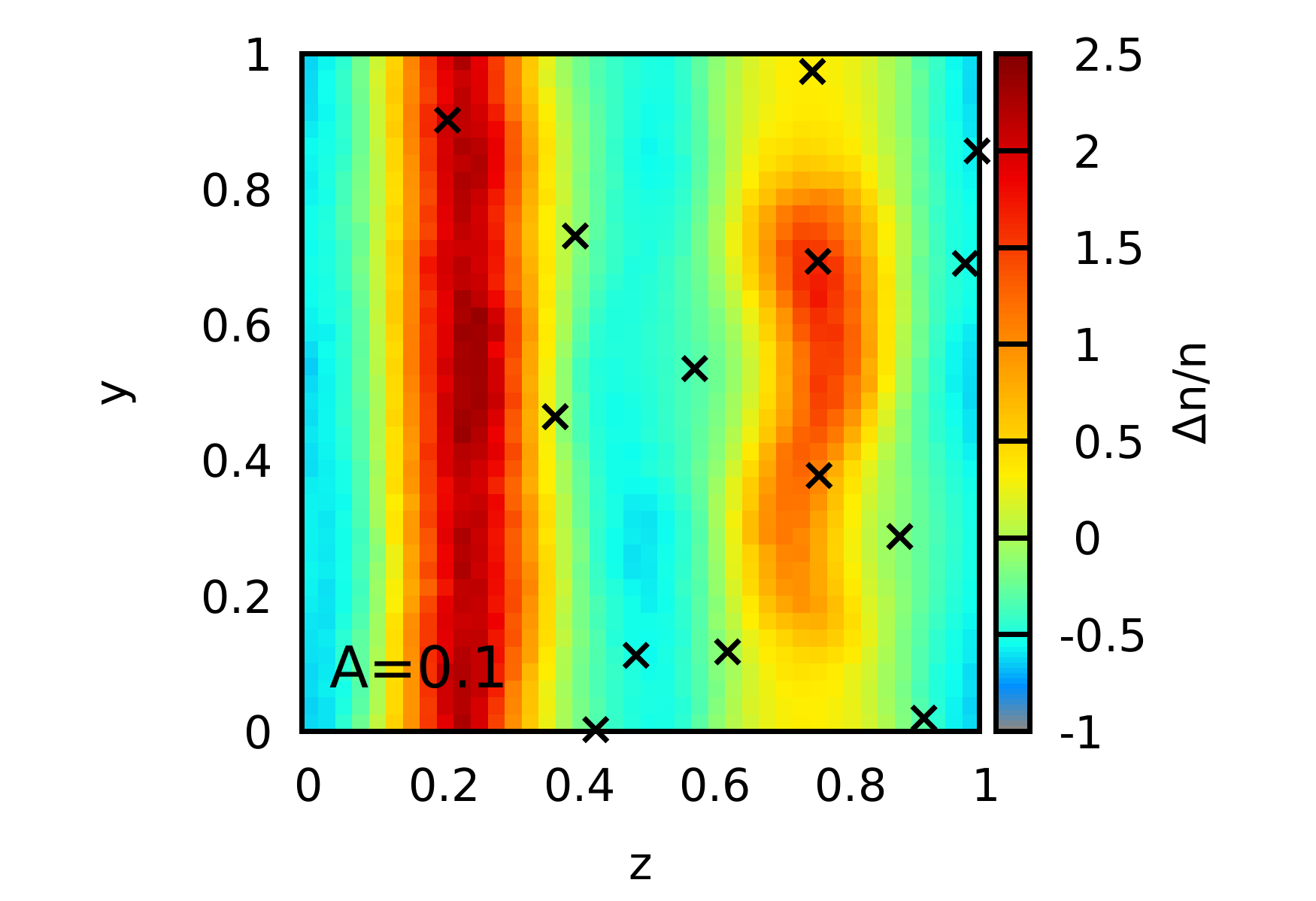}\\\hspace*{-0.5cm}\includegraphics[width=0.55\textwidth]{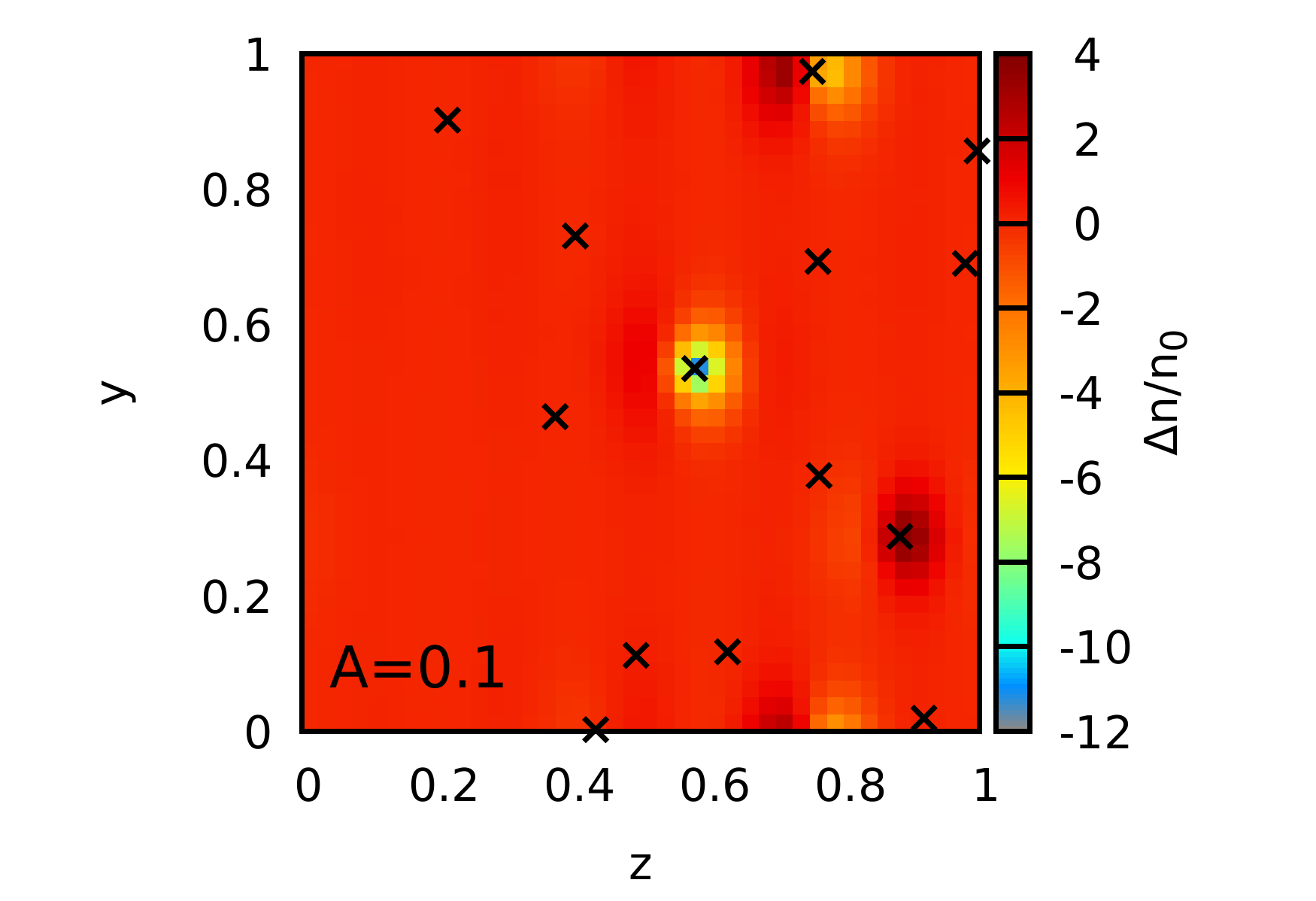}\hspace*{-1cm}\includegraphics[width=0.55\textwidth]{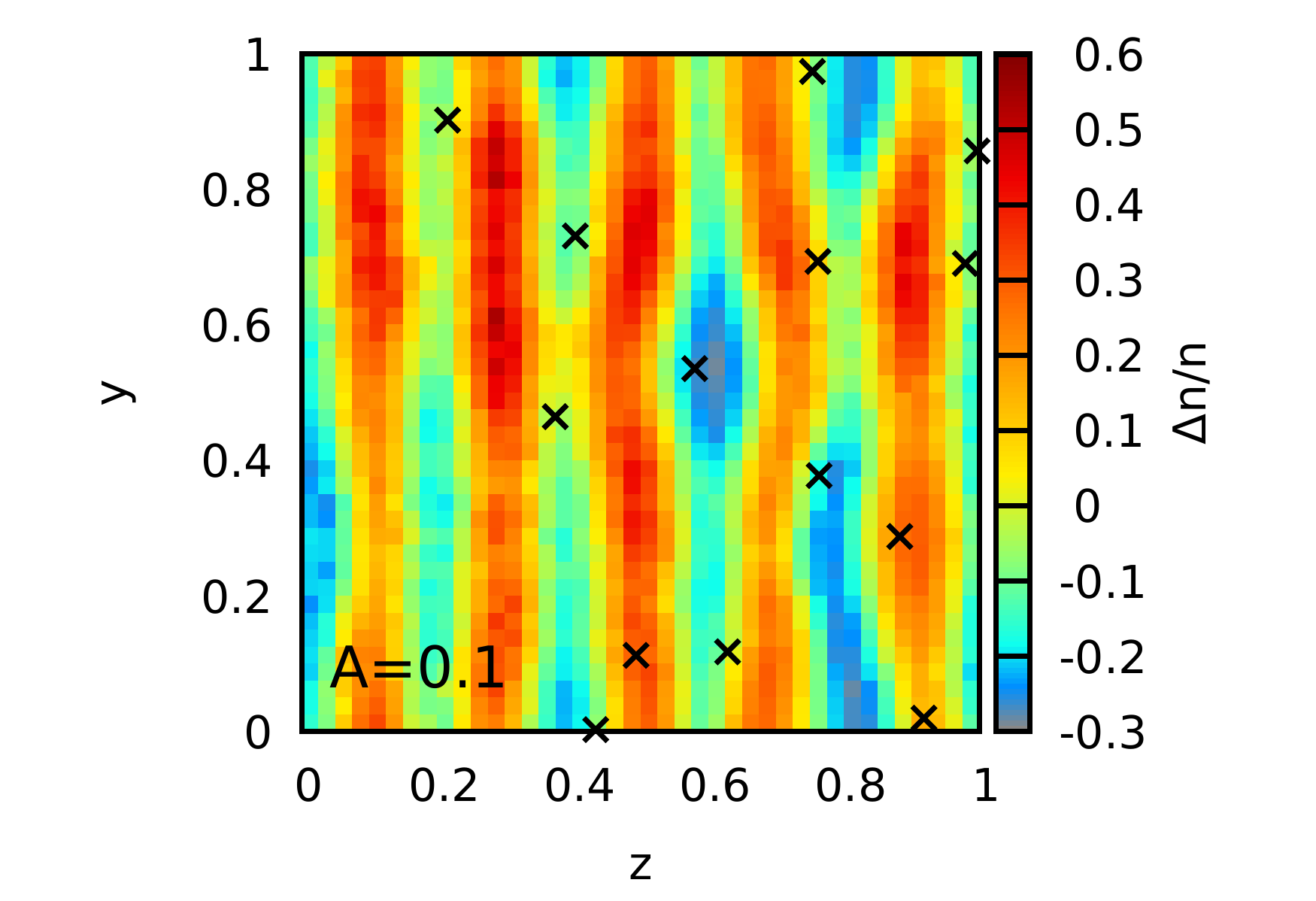}
\caption{\label{fig:panel}
PIMC results for the induced electronic density for $N=14$, $r_s=4$, and $\Theta=1$ for $A=0.1$. Top row: $\mathbf{q}=2\pi/L(0,0,1)^T$; center row: $\mathbf{q}=2\pi/L(0,0,2)^T$; bottom row: $\mathbf{q}=2\pi/L(0,0,5)^T$. The left and right columns show the change in the density in units of $n_0$, and the relative change in the density, respectively.
}
\end{figure*}

In Fig.~\ref{fig:panel}, we investigate in detail the response of this system to an external harmonic perturbation of amplitude $A=0.1$, which is close to, though somewhat beyond the linear-response regime~\cite{Bohme_PRL_2022}. More specifically, the top, center and bottom rows have been obtained for $\mathbf{q}=2\pi/L(0,0,1)^T$, $\mathbf{q}=2\pi/L(0,0,2)^T$, and $\mathbf{q}=2\pi/L(0,0,5)^T$, respectively, and the left and right columns show the density change (compared to the unperturbed system) in units of $n_0$, and the relative density change, i.e, $\Delta n_{x,y}(z)/n_{x,y}(z)$.
We observe two main trends, which are the same for all values of the wave vector $\mathbf{q}$. First, the absolute value of the density response is more pronounced in the vicinity of the ions; it positively correlates with the electronic density $n(\mathbf{r})$. Second, the relative density response exhibits the opposite trend, and is reduced around the protons. This holds both in regions where the cosinusoidal potential is positive (negative induced density) and negative (positive induced density).
In addition, we find that the induced change in the density is the largest for $\mathbf{q}=2\pi/L(0,0,2)^T$. This is expected, as the static density response function $\chi(\mathbf{q})$ attains a maximum modulus value for intermediate $q$~\cite{Bohme_PRL_2022}.

\begin{figure}\centering
\includegraphics[width=0.5\textwidth]{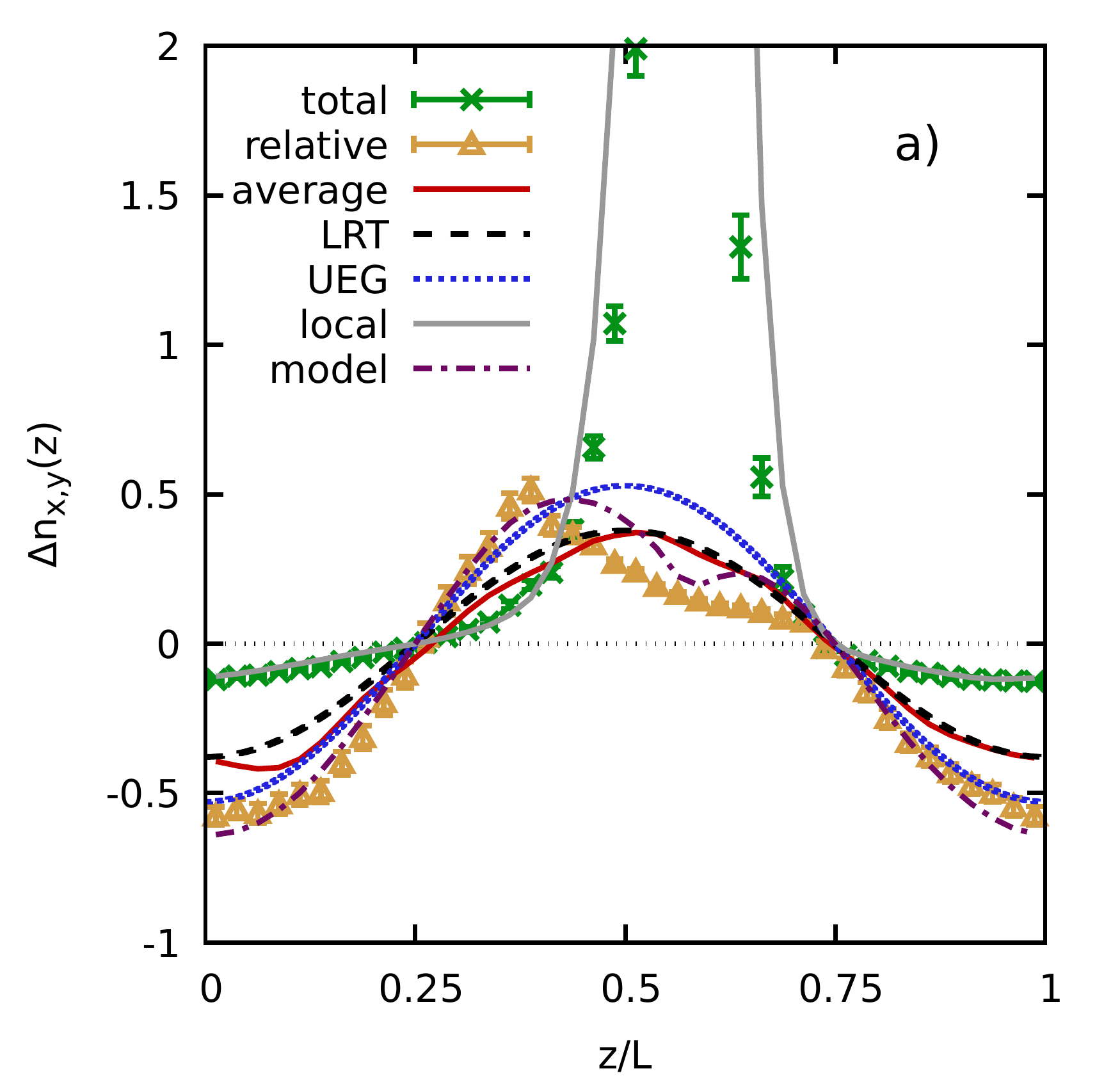}\\\vspace*{-1cm}
\includegraphics[width=0.5\textwidth]{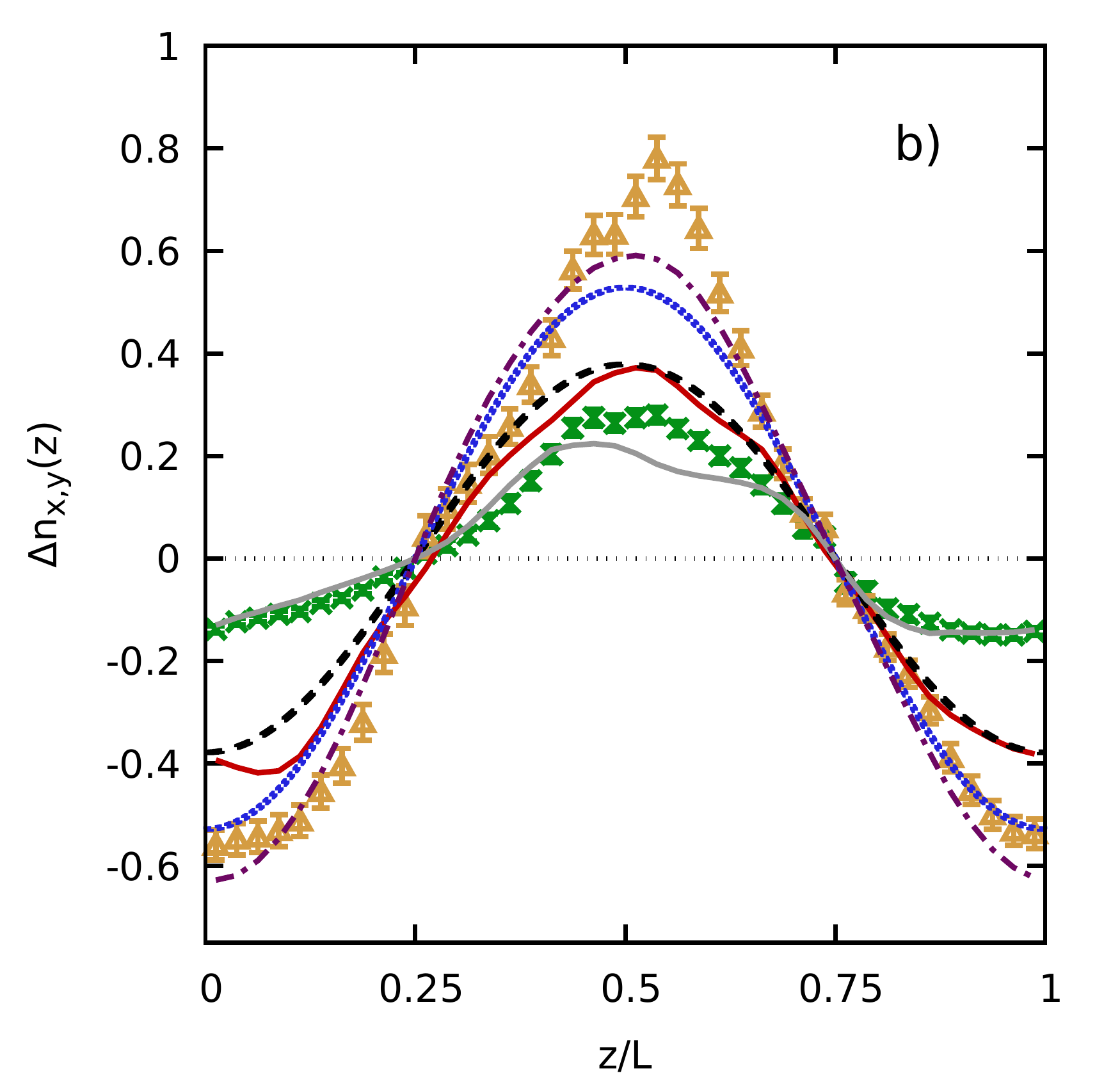}
\caption{\label{fig:N14_DELTA_SCAN}
Induced density change due to an external harmonic perturbation with $N=14$, $\mathbf{q}=2\pi/L(0,0,1)^T$, $A=0.1$, $P=500$. a) [b)]: scan line 1 [2]; see Fig.~\ref{fig:N14_rs4_theta1_unperturbed}. Green crosses: $\Delta n_{x,y}(z)$ in units of $n_0$; yellow triangles: $\Delta n_{x,y}(z)$ in units of $n_{x,y}(z)$; solid red: relative change in the density along the $z$-direction averaged over $x$ and $y$; dashed black: LRT fit; dotted blue: LRT result for the UEG; solid grey: local UEG model, Eq.~(\ref{eq:local}); dash-dotted purple: local density response model, Eq.~(\ref{eq:model}).
}
\end{figure}

To investigate these trends in more detail, we show the density change along two representative scan lines (cf.~the dashed red lines in Fig.~\ref{fig:N14_rs4_theta1_unperturbed}) in Fig.~\ref{fig:N14_DELTA_SCAN} for $\mathbf{q}=2\pi/L(0,0,1)^T$. The top panel corresponds to scan line 1, which crosses a proton at $z\approx 0.6L$; this can be easily seen in the green crosses depicting $\Delta n/n_0$. As noted above, most contributions to the induced density come from the vicinity of the proton where the density is maximal. In contrast, the relative change in the density (yellow triangles) exhibits the opposite trend.
In addition, we include the induced density along the $z$-direction that has been averaged with respect to $x$ and $y$ as the solid red curve. It closely resembles an inverted cosinusoidal curve, as it is predicted by linear-response theory for uniform systems, see Eq.~(\ref{eq:n}) above. To test this observation, we have performed a cosine fit, which the density response function $\chi(\mathbf{q})$ being the only free parameter. The resulting curve is shown as the dashed black line, and it is indeed in good agreement with the $x$-$y$-averaged data set. The small residual difference between the two curves can be interpreted as a finite-size effect, and vanishes if one averages over a sufficient number of snapshots~\cite{moldabekov2023averaging}.
This, however, is not the objective of the present work, where we intend to focus on the miscroscopic structure, instead of averaging it out. 
The dotted blue curve shows the results for Eq.~(\ref{eq:n}), using the linear-density response function of the UEG. The larger amplitude in this case nicely illustrates the, on average, reduced density response of the electrons as a consequence of the ions, even though it might be locally increased, both in the relative or the absolute sense. 
For completeness, we note that extensive quantum Monte Carlo calculations for the linear density response of the UEG are available in the literature~\cite{moroni,moroni2,cdop,dornheim_ML,dornheim_HEDP,dornheim_electron_liquid,Tolias_JCP_2021,Dornheim_HEDP_2022,Dornheim_review,Chen2019,PhysRevB.106.L081126,Dornheim_PRL_2020_ESA,Dornheim_PRB_ESA_2021}.

To get additional insights into the density response of hydrogen, we compare our simulation data with two heuristic models. Let us assume that the system responds to the external perturbation like a UEG, but re-scaled by the inhomogeneous density profile of the unperturbed system, $n(\mathbf{r})$. This leads to the density profile
\begin{eqnarray}\label{eq:local}
   \Delta n_\textnormal{local}(\mathbf{r}) = n_0 + 2A\ \chi_\textnormal{UEG}(\mathbf{q})\ \textnormal{cos}\left(\mathbf{q}\cdot\mathbf{r}\right)\ \frac{n(\mathbf{r})}{n_0} \ .
\end{eqnarray}
The results for $\Delta n_\textnormal{local}(\mathbf{r}) /n_0$ are included into Fig.~\ref{fig:N14_DELTA_SCAN} as the solid light grey curve, which qualitatively, though not quantitatively reproduces the green crosses. Specifically, Eq.~(\ref{eq:local}) overestimates the actual density response around the protons.
As a second model, we drop the weighting factor of $n(\mathbf{r})/n_0$ from Eq.~(\ref{eq:local}) and instead make the density response function of the UEG depend on the local value of the density, 
\begin{eqnarray}\label{eq:model}
   \Delta n_\textnormal{model}(\mathbf{r}) = n_0 + 2A\ \chi_\textnormal{UEG}\left[\mathbf{q};n(\mathbf{r})\right]\ \textnormal{cos}\left(\mathbf{q}\cdot\mathbf{r}\right)\  \ .
\end{eqnarray}
This leads to the dash-dotted purple curve in Fig.~\ref{fig:N14_DELTA_SCAN}, which is on average close to the dotted blue curve representing a pure UEG, but also includes some local structure resembling the yellow triangles. This can be understood by recalling the exact long-wavelength limit of the static linear UEG density response function, which is given by~\cite{kugler_bounds}
\begin{eqnarray}\label{eq:limit}
    \lim_{q\to0}\chi_\textnormal{UEG}(q) = - \frac{q^2}{4\pi}\ ,
\end{eqnarray}
with $q=|\mathbf{q}|$. Since Eq.~(\ref{eq:limit}) holds independent of the density, the local density response function $\chi_\textnormal{UEG}\left[\mathbf{q};n(\mathbf{r})\right]$ in Eq.~(\ref{eq:model})
only weakly depends on the density for the comparably small value of $q$ considered here. This explains its similarity to the pure UEG curve, rather than to the strongly inhomogeneous absolute response. 


In Fig.~\ref{fig:N14_DELTA_SCAN}b), we repeat this analysis for scan line 2, which is located in a region without protons, and, therefore, with a lower density that does not exhibit large density gradients. We note that the solid red, dashed black, and dotted blue curves are the same as in panel a) and have been included as a reference. As it is expected, the absolute induced density (in units of $n_0$, green crosses) is smaller by more than an order of magnitude compared to the first scan line. It is qualitatively well reproduced by the re-scaled UEG model defined in Eq.~(\ref{eq:local}), which is a consequence of the small variations in the density. Furthermore, the local density response model from Eq.~(\ref{eq:model}) again closely agrees with the pure UEG curve for this value of $q$. Finally, we observe that the relative change in the density computed from our PIMC simulations (yellow triangles) overall exceeds the response of the UEG in magnitude. Taken together, panels a) and b) thus reveal the following: 1) the density response of the hydrogen snapshot is strongly inhomogeneous and qualitatively follows the unperturbed density profile $n(\mathbf{r})$ [cf.~Eq.~(\ref{eq:local})]; 2) the relative response is increased in low-density regions and decreased in high-density regions (i.e., around protons) compared to the UEG; on average, the decrease predominates over the increase, leading to an overall reduction in the density response of hydrogen compared to the UEG~\cite{Bohme_PRL_2022}; 4) the decomposition into effectively \emph{bound} and \emph{free} electrons, with the latter resembling the behaviour of a UEG model, is questionable. In particular, all electrons in the region investigated in 
Fig.~\ref{fig:N14_DELTA_SCAN}b) would have to be considered as \emph{free}, but their density response differs significantly from the UEG.

\begin{figure}\centering
\includegraphics[width=0.5\textwidth]{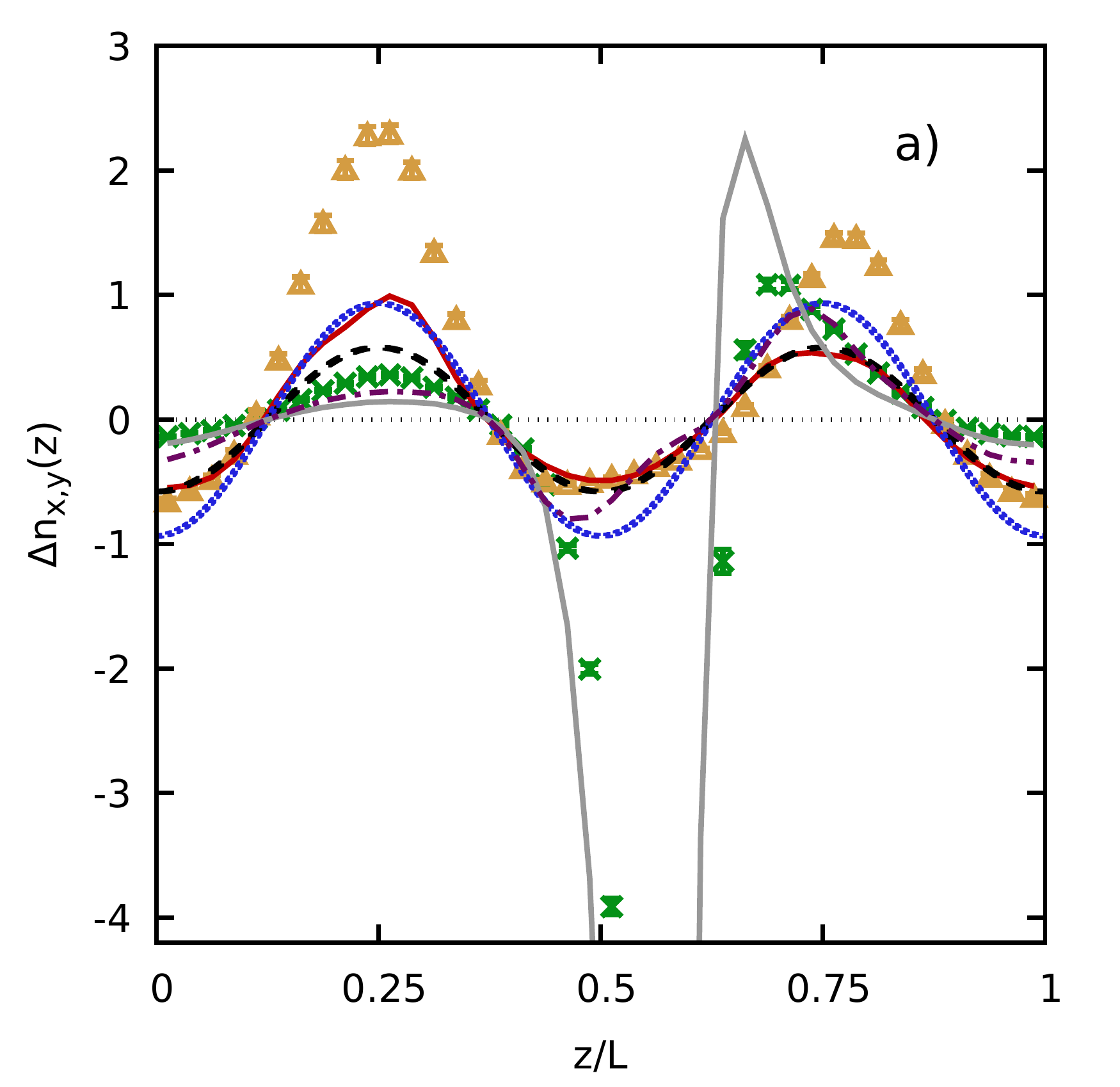}\\\vspace*{-1cm}\hspace*{-0.032\textwidth}
\includegraphics[width=0.53\textwidth]{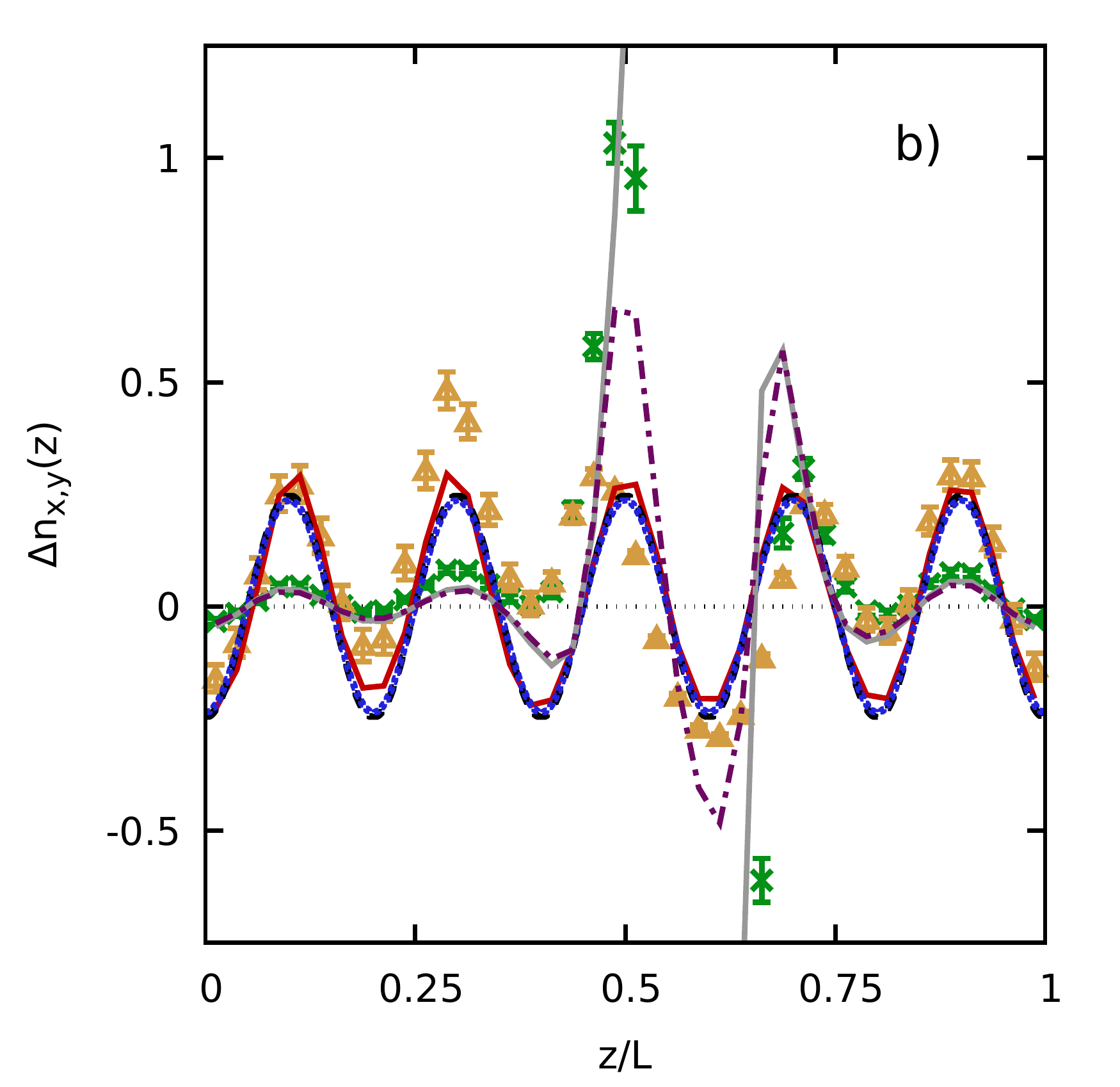}
\caption{\label{fig:N14_DELTA_SCAN_large_q}
Induced density change due to an external harmonic perturbation with $N=14$, and $A=0.1$, $P=500$. a)
$\mathbf{q}=2\pi/L(0,0,2)^T$; b) $\mathbf{q}=2\pi/L(0,0,5)^T$. Green crosses: $\Delta n_{x,y}(z)$ in units of $n_0$; yellow triangles: $\Delta n_{x,y}(z)$ in units of $n_{x,y}(z)$; solid red: relative change in the density along the $z$-direction averaged of $x$ and $y$; dashed black: LRT fit; solid grey: local UEG model, Eq.~(\ref{eq:local}); dash-dotted purple: local density response model, Eq.~(\ref{eq:model}).
}
\end{figure}

In Fig.~\ref{fig:N14_DELTA_SCAN_large_q}, we extend this analysis to the larger wave vectors $\mathbf{q}=2\pi/L(0,0,2)^T$ (a) and $\mathbf{q}=2\pi/L(0,0,5)^T$ (b). Generally, the PIMC results for the absolute change in the density follow the same trend as in Fig.~\ref{fig:N14_DELTA_SCAN}. In addition, they are qualitatively reproduced, but generally somewhat overestimated, by the grey curve computed from the locally density-weighted UEG model, Eq.~(\ref{eq:local}), for both values of $\mathbf{q}$. Regarding the relative change in the density (yellow triangles), we find the largest difference to the $x$-$y$-averaged curve (solid red) for $\mathbf{q}=2\pi/L(0,0,2)^T$. This can be understood intuitively by considering the involved length scales. Specifically, a cosinusoidal perturbation of wave vector $\mathbf{q}$ is associated with the wave length $\lambda=2\pi/q$. For large $q$, $\lambda$ is substantially smaller than the average interparticle distance, and the density response is thus dominated by the single-electron limit~\cite{Dornheim_Nature_2022,Dornheim_review}; the latter is comparably insensitive to the local ionic structure. Indeed, B\"ohme \emph{et al.}~\cite{Bohme_PRL_2022} have found close agreement between the density response of hydrogen and the UEG model for $q\gtrsim4q_\textnormal{F}$, and the close agreement between the dashed black and dotted blue curves in panel b) further corroborate this observation.
For the $\mathbf{q}$ value investigated in Fig.~\ref{fig:N14_DELTA_SCAN_large_q}a), on the other hand, the wavelength is comparable to the average interparticle distance, which means that the sensitivity to the local structure is most pronounced. This explains 1) the observed large difference between the yellow triangles and the $x$-$y$-averaged (solid red) curve  and 2) the comparably larger deviation between the averaged hydrogen results and the pure UEG model.

A further interesting observation is the dependence of the local density response function model [Eq.~(\ref{eq:model})] on the wave vector $\mathbf{q}$. While it was not able to reproduce the large local variations in the induced density profile in the long wavelength limit, this situation substantially changes with increasing $q$. For $\mathbf{q}=2\pi/L(0,0,2)^T$, the corresponding purple dash-dotted curve strongly deviates from the purely cosinusoidal UEG model and qualitatively reproduces the PIMC results for the absolute change in the density (green crosses), except in the nearest vicinity of the proton.  
For $\mathbf{q}=2\pi/L(0,0,5)^T$, the deviations from the UEG are even more pronounced and again reproduce the green crosses, except for $z\approx0.6 L$.

\subsection{Dependence on the density\label{sec:density}}

\begin{figure}\centering
\hspace*{-0.5cm}\includegraphics[width=0.55\textwidth]{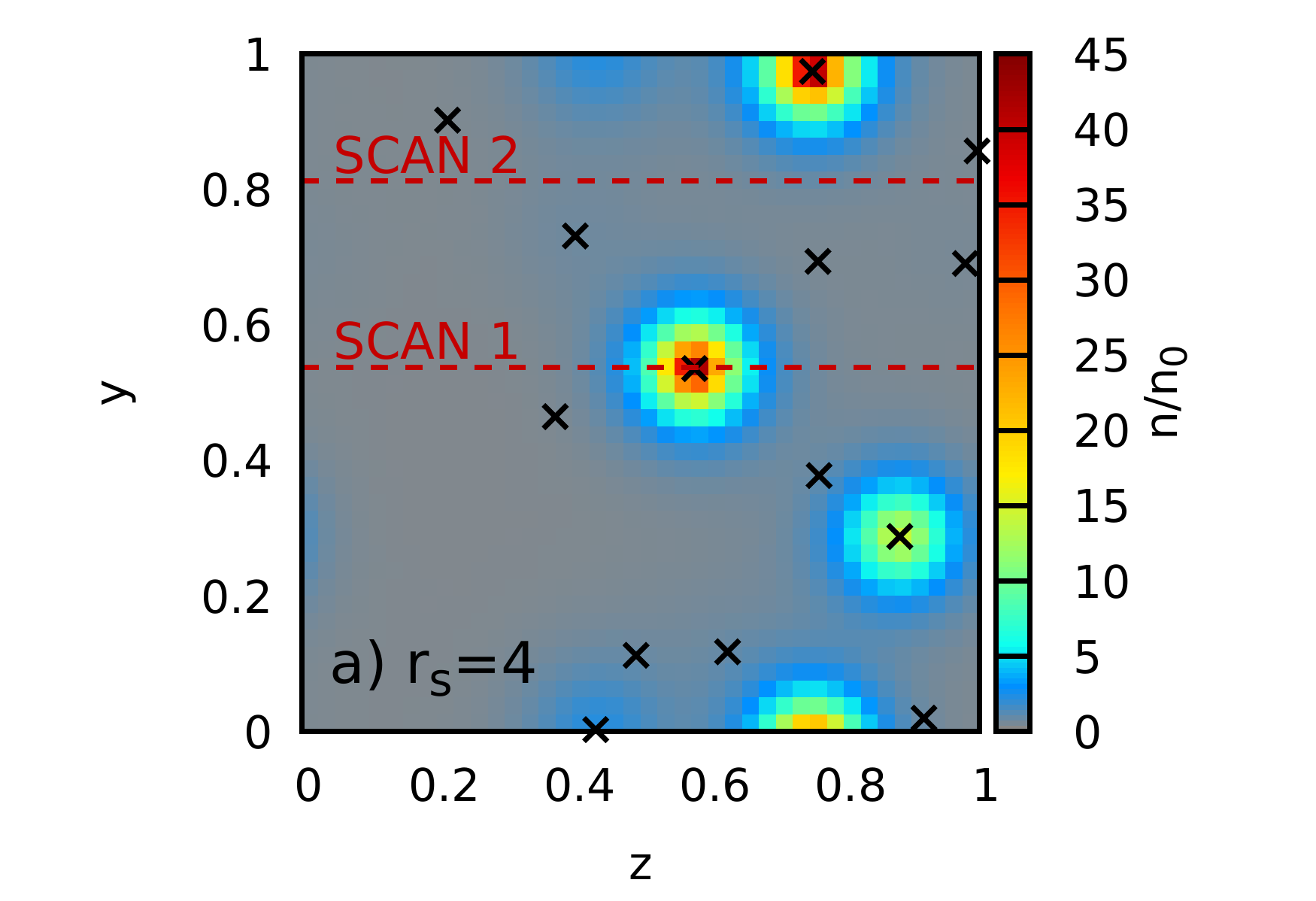}\\\vspace*{-0.85cm}
\hspace*{-0.5cm}\includegraphics[width=0.55\textwidth]{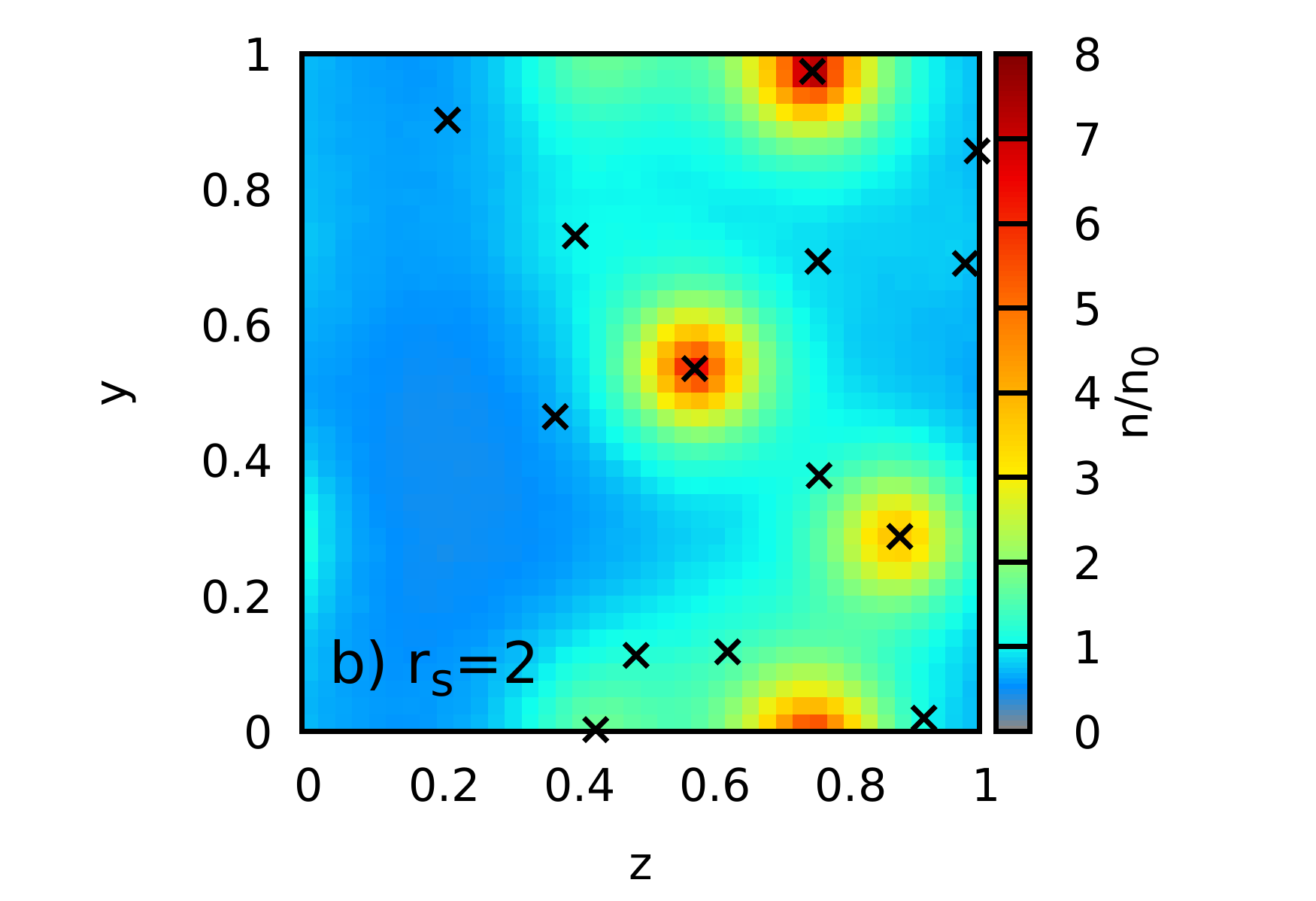}\\\vspace*{-0.85cm}
\hspace*{-0.5cm}\includegraphics[width=0.55\textwidth]{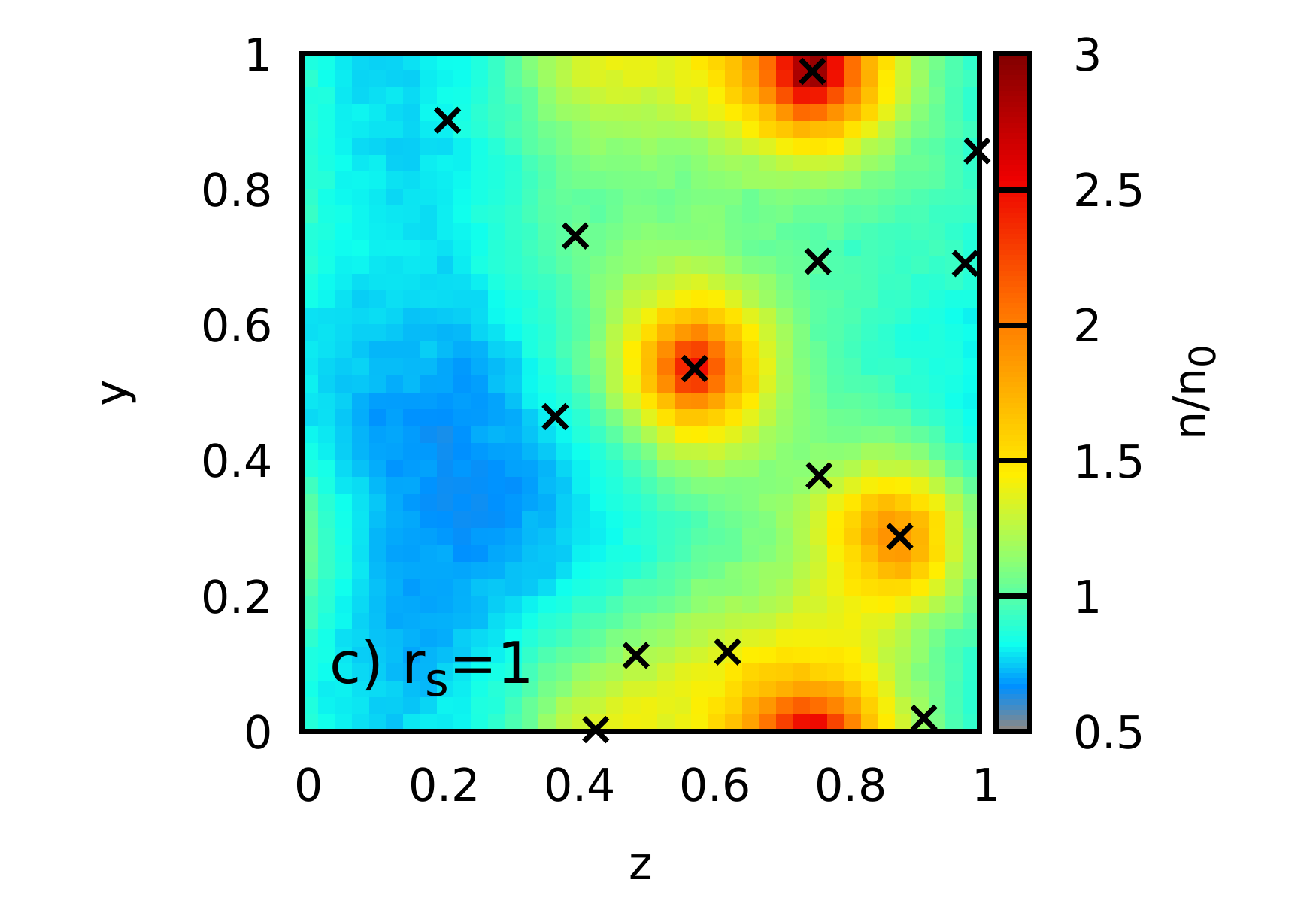}
\caption{\label{fig:C2P_rs}
PIMC results for the electronic density in the $z$-$y$-plane (cf.~the yellow surface in Fig.~\ref{fig:Snap_N14_rs4_theta1}) for $N=14$, $\Theta=1$, and $P=500$. a) $r_s=4$; b) $r_s=2$; c) $r_s=1$.
}
\end{figure} 

A further important question is the interplay between the ionic structure, the electronic density response, and the density parameter $r_s$. In Fig.~\ref{fig:C2P_rs}, we show PIMC results for the density in the $x$-$y$-plane (cf.~the yellow surface in Fig.~\ref{fig:Snap_N14_rs4_theta1}) for the same configuration of protons and $r_s=4$ (top), $r_s=2$ (center), and $r_s=1$ (bottom). To isolate the impact of the density, we keep the degeneracy temperature, rather than $T$ itself, constant as $\Theta=1$. From a physical perspective, $r_s=2$ corresponds to a metallic density, and $r_s=1$ constitutes a strongly compressed state that can be probed for example at ICF experiments at the National Ignition Facility~\cite{Moses_NIF}.

With decreasing the Wigner-Seitz radius (i.e., increasing the average electronic number density), the electrons become substantially less localized. The electron density is increased by a factor of less than three around the protons for $r_s=1$, compared to the forty-fold increase at $r_s=4$. Indeed, B\"ohme \emph{et al.}~\cite{Bohme_PRL_2022} have reported that the density response of hydrogen strongly resembles the behaviour of a UEG for $r_s=2$ and $\Theta=1$.

\begin{figure}\centering
\includegraphics[width=0.5\textwidth]{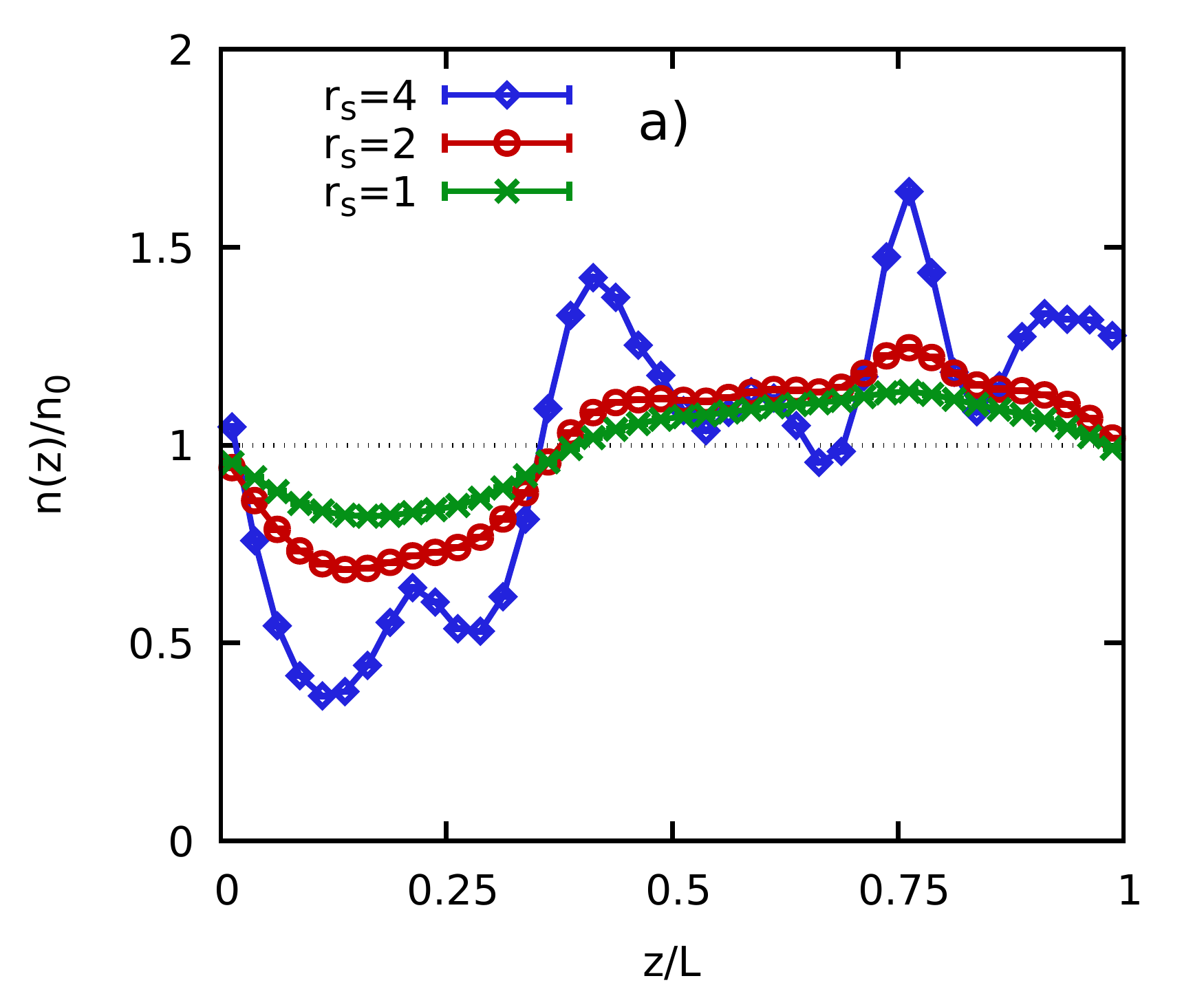}\\\vspace*{-1cm}
\hspace*{0.0125\textwidth}\includegraphics[width=0.486\textwidth]{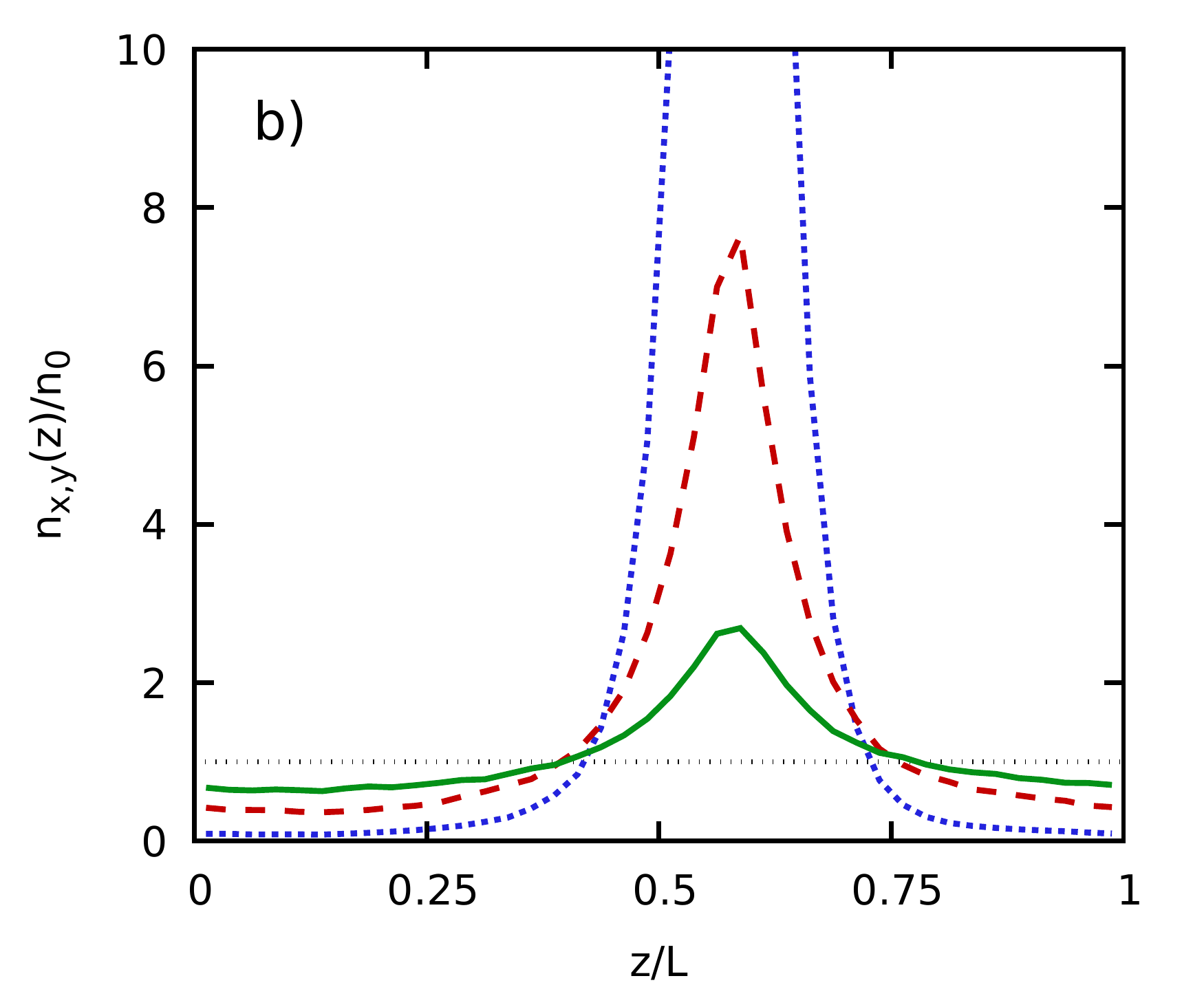}\\\vspace*{-1cm}
\includegraphics[width=0.5\textwidth]{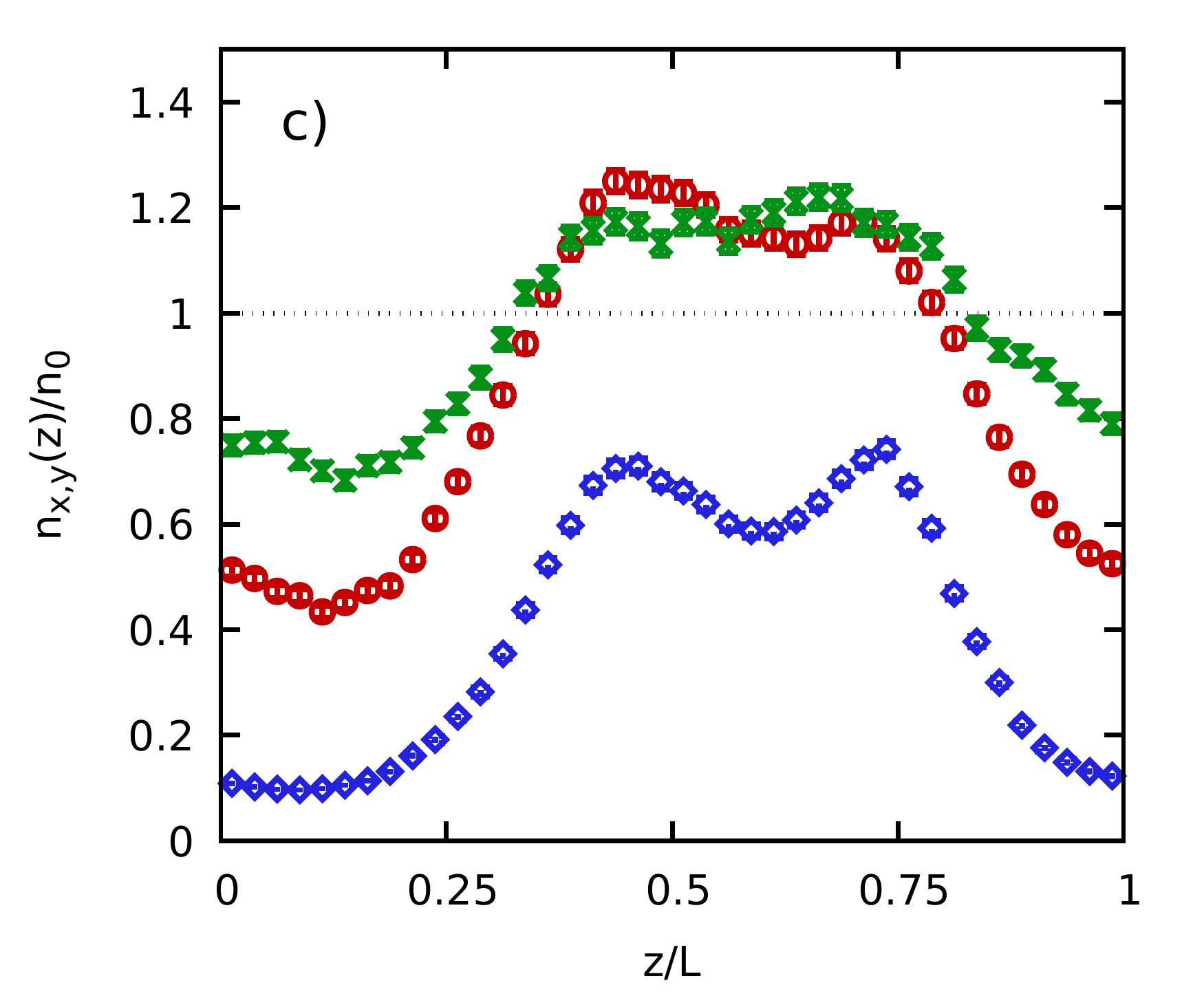}
\caption{\label{fig:N14_rs_strips}
a) $x$-$y$-averaged density along the $z$-direction; b) [c)]: density along scan line 1 [scan line 2], cf.~Fig.~\ref{fig:C2P_rs}, without external perturbation.
}
\end{figure}

This can also be seen very clearly in Fig.~\ref{fig:N14_rs_strips}a), where we show the $x$-$y$-averaged density along the $z$-direction for the three considered values of the density. Only the PIMC results for $r_s=4$ (blue diamonds) exhibit a pronounced structure, whereas the data sets for $r_s=2$ (red circles) and in particular $r_s=1$ (green crosses) are comparably featureless. 
In Figs.~\ref{fig:N14_rs_strips}b) and c), we show scan lines over the $x$-$y$-plane (cf.~the dashed red lines in Fig.~\ref{fig:C2P_rs}). Panel b) corresponds to scan line 1 that includes a proton at $z\approx0.6L$; it again nicely illustrates the drastic decrease of the electronic localization with increasing density in this regime. Panel c) corresponds to a region without a proton, such that the density is depleted for $r_s=4$ over the entire $z$-range. This is not true for $r_s=2$ and $r_s=1$, for which $n_{x,y}(z)$ fluctuates around the average value of $n_0$.

\begin{figure}\centering
\includegraphics[width=0.5\textwidth]{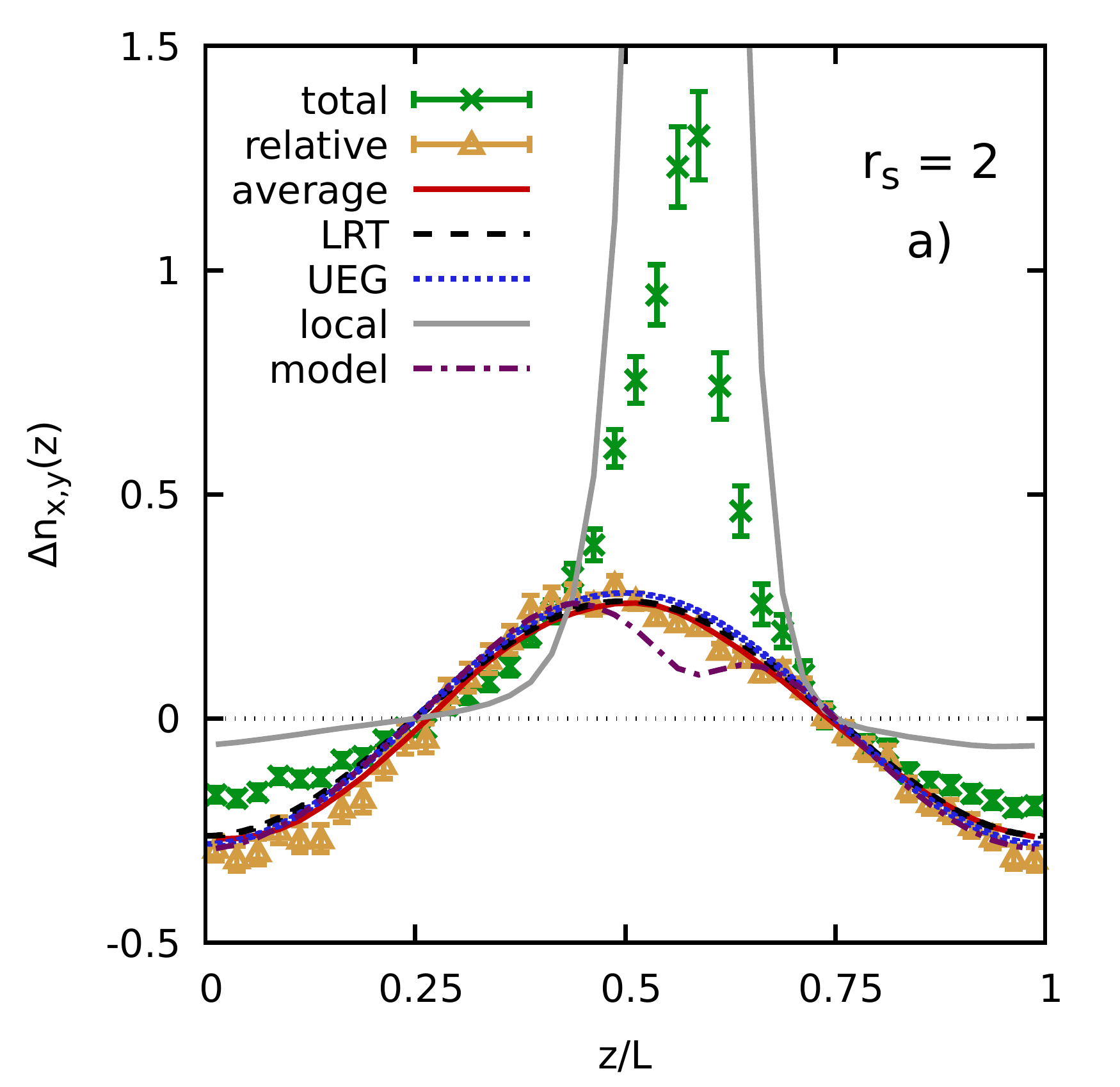}\\\vspace*{-1cm}
\includegraphics[width=0.5\textwidth]{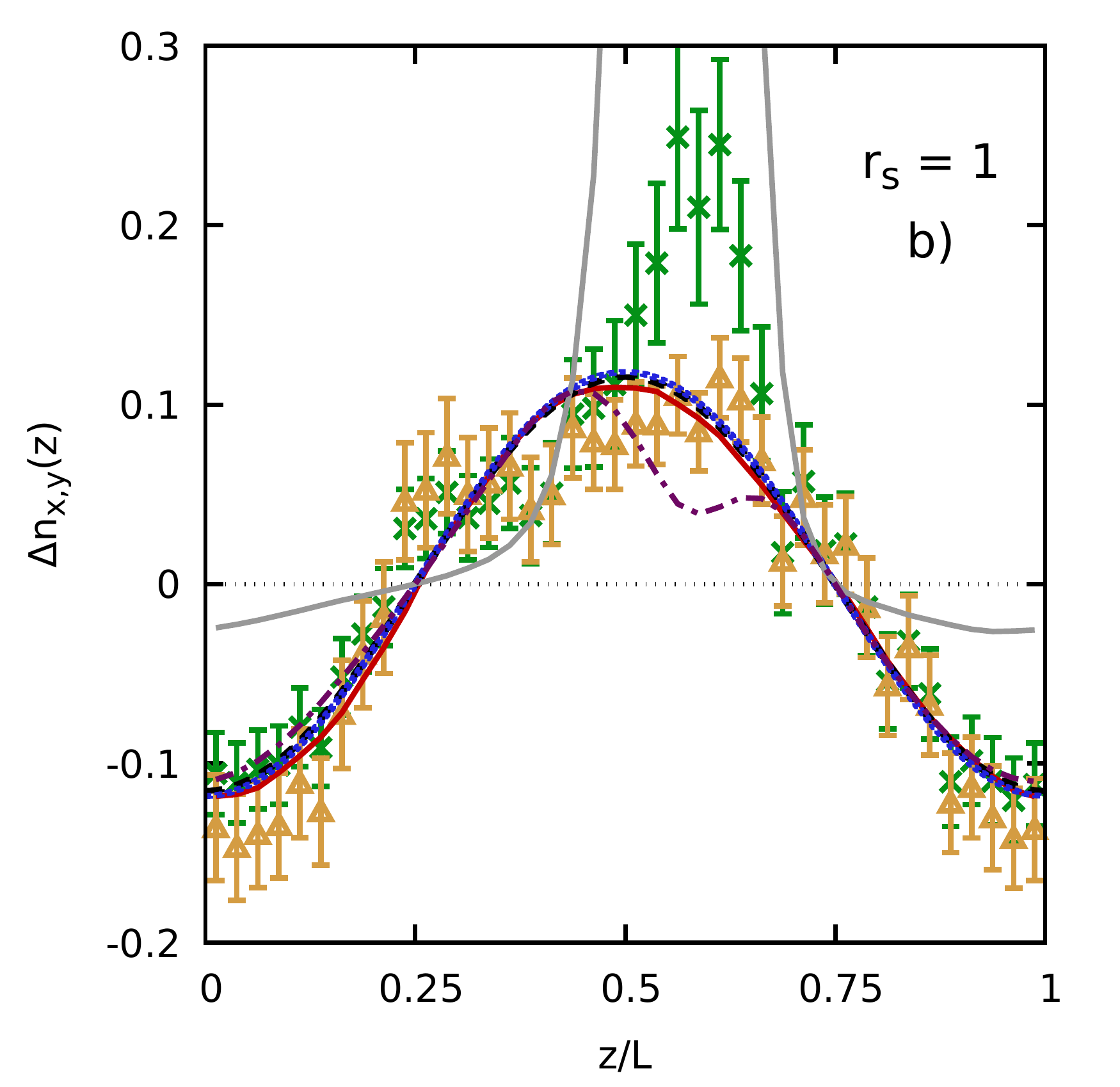}
\caption{\label{fig:N14_rs}
Induced density change due to an external harmonic perturbation with $N=14$, $\mathbf{q}=2\pi/L(0,0,1)^T$, and $P=500$. a) $r_s=2$ ($A=0.15$); b) $r_s=1$ ($A=0.2$). Green crosses: $\Delta n_{x,y}(z)$ in units of $n_0$; yellow triangles: $\Delta n_{x,y}(z)$ in units of $n_{x,y}(z)$; solid red: relative change in the density along the $z$-direction averaged of $x$ and $y$; dashed black: LRT fit; solid grey: local UEG model, Eq.~(\ref{eq:local}); dash-dotted purple: local density response model, Eq.~(\ref{eq:model}).
}
\end{figure}

Let us conclude this analysis by considering the induced density along scan line 1 due to an external harmonic perturbation of wave vector $\mathbf{q}=2\pi/L(0,0,1)^T$ shown in Fig.~\ref{fig:N14_rs} for $r_s=2$ (top) and $r_s=1$ (bottom). The most important trend is that the absolute response (green crosses) increasingly resembles the relative response (yellow triangles); this is a direct consequence of $n(\mathbf{r})\to n_0$ with increasing density. Furthermore, all data sets increasingly resemble the UEG model for the same reason. In fact, even the absolute density response is almost undistinguishable from the UEG model at $r_s=1$.

\section{Summary and Outlook\label{sec:summary}}

In this work, we have presented extensive new \emph{ab initio} PIMC results for the electronic density response of warm dense hydrogen on the nanoscale. This has been achieved by taking a snapshot with fixed proton positions from a DFT-MD simulation, and solving the electronic problem in this external potential with PIMC.
We note that we use the direct PIMC method without any restrictions on the nodal surface of the density matrix, which makes our simulations computationally expensive, but exact within the given Monte Carlo error bars.

To demonstrate the quality of our simulations, we have carried out an in-depth analysis of the convergence with the number of high-temperature factors $P$. By utilizing the exact solution for the thermal density matrix of the electron-ion two-body problem (\emph{pair approximation}), we attain an accuracy of $\sim0.1\%$ with $P=500$ high-temperature factors even in the direct vicinity of the protons, where the density gradients are most pronounced.
The diagonal Kelbg potential, too, converges towards the exact result with increasing $P$, albeit substantially slower; this is consistent with previous investigations~\cite{Bohme_PRE_2023,Filinov_PRE_2004}.

From a physical perspective, our study has given us important insights into two important trends. First, we have studied in detail the localization of the electrons around the protons on the nanoscale, without any assumptions about \emph{bound} or \emph{free} electrons. For $r_s=4$, the electrons are strongly localized, whereas hydrogen more closely resembles the well-known UEG model for $r_s=2$ and especially for $r_s=1$. 
Second, we have studied the interplay between the electronic density response to an external static harmonic perturbation and the presence of the protons. Here, our main findings include the fact that the absolute density response positively correlates with the electron density $n(\mathbf{r})$, whereas the opposite holds for the relative density response. Moreover, the spatially resolved density response of hydrogen can be modelled qualitatively by the behaviour of a local UEG model, but this assumption does not capture the reduction of the averaged density response compared to a pure UEG due to the presence of the protons.

Our findings are of direct consequence for upcoming experiments at different facilities. For example, modelling XRTS measurements of hydrogen jets~\cite{Zastrau} is likely challenging due to the expected low densities; it will require to accurately capture the complicated interplay between the electrons and the ions, and to incorporate the strong degree of electronic localization.
In contrast, ICF experiments with hydrogen isotopes at the National Ignition Facility might potentially already be reproduced by a UEG model, although this has to be carefully checked in practice.

In addition to being interesting in their own right, we expect our setup to be of high value for the benchmarking of less accurate methods for the simulation of WDM. This includes, but is not limited to, a rigorous assessment of commonly used exchange--correlation functionals for DFT, and the fixed-node approximation in restricted PIMC simulations.
Other future works will include full PIMC simulations of warm dense hydrogen, where, instead of being kept fixed, the ions are treated on the same level of the electrons. This will allow for extensive studies of a gamut of density response properties, including the linear and nonlinear regimes, as well as dynamic properties based either on an analytic continuation~\cite{dornheim_dynamic,dynamic_folgepaper,JARRELL1996133} or directly in the imaginary-time domain~\cite{Dornheim_insight_2022,Dornheim_review,Dornheim_PTR_2022,dornheim2023extraction}.

\section*{Acknowledgments}
This work was partially supported by the Center for Advanced Systems Understanding (CASUS) which is financed by Germany’s Federal Ministry of Education and Research (BMBF) and by the Saxon state government out of the State budget approved by the Saxon State Parliament. This work has received funding from the European Research Council (ERC) under the European Union’s Horizon 2022 research and innovation programme
(Grant agreement No. 101076233, "PREXTREME").
The PIMC calculations were partly carried out at the Norddeutscher Verbund f\"ur Hoch- und H\"ochstleistungsrechnen (HLRN) under grant shp00026 and on a Bull Cluster at the Center for Information Services and High Performance Computing (ZIH) at Technische Universit\"at Dresden.

\bibliography{bibliography}
\end{document}